# A New View of the Universe

Ben Kristoffen

c/o K. Krogh, Neuroscience Research Institute
University of California, Santa Barbara, CA  93106
e-mail: krogh@psych.ucsb.edu *

**Abstract:**  A new, very different physical model of the universe is proposed.  Its virtues include unifying relativity and quantum mechanics, and particles with de Broglie waves.  It also appears to provide a truly unified physical basis for electromagnetic, gravitational and nuclear forces.  The basic system is a four-dimensional Euclidean space, containing an array of nonlinear "flow" waves.  These repeat in one dimension, called $\phi$.  As in Newtonian mechanics, time is treated as an additional, unidirectional parameter describing the evolution of the system.  Nevertheless, this wave system is shown be inherently relativistic.  Further self-organizing patterns arise within the overall wave structure.  Called "wavicles," these have *intrinsic* quantized fields, "spin," and rest energy, and represent elementary particles.  Relativistic expressions are derived for particle behavior in scalar, vector and gravitational potentials.  Proper representations of these potentials, based on the wave fields and associated $\phi$ flows of wavicles, are also obtained.  As in the causal quantum mechanics of de Broglie and Bohm, wavicles exist continuously and follow definite, stochastic trajectories.  *Although experimentally equivalent to Einstein's special relativity, this theory differs fundamentally from his general relativity and the associated Big Bang model.*  According to Linde, the latter predict a large-scale space-time curvature roughly 60 orders of magnitude greater than observed values.  Here a flat large-scale geometry is predicted, in agreement with the observed distribution of galaxies.  This theory is also consistent with recent observations pertaining to the age of the universe.

# Contents

# 1. Introduction

> ...a good theoretical physicist today might find it useful to have a wide range of physical viewpoints and mathematical expressions of the same theory (for example of quantum electrodynamics) available to him . . . If [everyone] follows the same current fashion in expressing and thinking about electrodynamics or field theory, then the variety of hypotheses being generated . . . is limited. Perhaps rightly so, for possibly the chance is high that the truth lies in the fashionable direction. But, on the off chance that it is in another direction -- a direction obvious from an unfashionable view of field theory -- who will find it? Only someone who has sacrificed himself by teaching himself quantum electrodynamics from a peculiar and unusual point of view, one he may have to invent for himself.
> - Richard Feynman, Nobel Prize lecture

What are the constituents of matter? Although they are equally particle and wave-like, we call them particles. Early in the development of quantum mechanics, Erwin Schrödinger expressed the view that it should be possible to explain particles on the basis of wave phenomena, and thereby provide a unified representation of both particle aspects. However, Schrödinger was unable to do so, and while there have been other notable efforts, by de Broglie[1], Bohm[2], and Vigier[3], for example, or Pering and Skyrme[4], none have found widespread acceptance.

In widely accepted theories, such as quantum chromodynamics, the complexity and richness of particles and their interactions are attributed to their particle-like aspect. The associated matter waves are assumed to be relatively featureless and play no more than the simple probabilistic role first articulated by Born in the mid-nineteen twenties. Explored here is a new theory, related to that of de Broglie, Bohm and Vigier, where waves play a much broader role. Besides a unified picture of matter waves and particles, it appears to offer a unified description of relativity, quantum mechanics, electrodynamics, gravitation, and nuclear forces as well.

Unlike the waves of standard quantum mechanics, the waves in this theory are nonlinear. Involved is a nonlinear effect, common in nature, whose importance is not widely appreciated. The waves are also highly organized, in what is sometimes called a coherent structure. Within this overall structure, additional self-organizing wave patterns arise: these correspond to both particles and their fields. Before taking up the new theory, I'd like to begin with a qualitative look at various known nonlinear wave phenomena and some coherent structures.

# Nonlinear Waves

Stanislaw Ulam remarked once that using the term "nonlinear science" is like calling zoology "the study of non-elephant animals." The same could be said of "nonlinear waves." Real waves of every kind have at least some small degree of nonlinearity. Even electromagnetic waves, often cited in textbooks as examples of perfect linearness, are not. (Nonlinearity in electromagnetic waves can give rise to electron-positron pairs, for example.) So I'd like to begin by asserting that, when we speak of "nonlinear waves," we are really talking about waves in general.

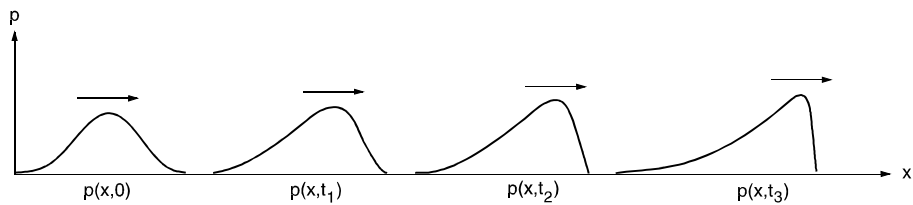

**Fig. 1-1.** Shock front development in an acoustic wave. (From Towne[5].)

A basic attribute of nonlinear waves, essential to the present theory, is feedback. A simple example can be seen in acoustic shock waves. In these, the propagation velocity depends markedly on wave amplitude -- the larger the amplitude, the greater the velocity. Fig. 1-1 depicts the development of a shock front. (Here the wave is propagating in just one dimension.) As the faster-moving peak overtakes the front of the wave, the more it grows. And the more it grows, the faster it overtakes the front, steepening the wavefront. So here there is positive feedback.

Ripples in wind-blown sand are a good example of negative feedback. This gives the ripples in a given region similar heights and spacings. (One obvious factor is that the wind blows the exposed tops off of taller ripples.) Sand ripples also have a capacity to counteract disturbances. If you rub out some of the ripples in an area, with a steady wind, the pattern will "heal" itself. Feedback is, of course, an essential element in many self-organizing natural systems, the most remarkable being life. While sand ripples exhibit only simple behaviors, there are other wave systems which may seem almost alive.

Such waves are found in the plasmas of fusion machines. These highly nonlinear waves, or "instabilities," have a remarkably powerful and adaptable capacity for self-organization. The plasmas are contained in magnetic "wells" with energy depths far greater than the overall plasma energies. However, the instabilities have high localized energy concentrations, which have invariably allowed them and their plasmas to escape the best designed wells.

Like insects evolving to defeat pesticides, as wells have been redesigned to defeat known instabilities, new types have always arisen to take their places.

What causes nonlinearity in waves? The obvious answer is: nonlinearity in the wave media. However, even with a perfectly linear medium (if there were such) nonlinearity can still arise. Nonlinearity in waves has a second cause, equally as important as the characteristics of the medium: the waves themselves. This aspect of wave behavior receives little attention in most university physics curricula. In elementary texts, the mechanism behind this nonlinearity is usually denied:

> If you have ever watched ocean waves moving toward shore, you may have wondered if the waves were carrying water into the beach. This is, in fact, not the case. Water waves move with a recognizable velocity. But each particle of water itself merely oscillates about an equilibrium point . . . This is a general feature of waves: waves can move over large distances, but the medium itself has only limited movement . . . A wave consists of oscillations that move without carrying matter with them.
> -Douglas C. Giancoli
> *Physics*, 3rd Ed., 1991

> *Waves transport energy, but not matter, from one region to another*. [Original italics.]
> -Hugh D. Young
> *University Physics*, 8th Ed., 1992

> ...here we see an essential feature of what is called wave motion. A condition of some kind is transmitted from one place to another by means of a medium, but the medium itself is not transported.
> -A. P. French
> *Vibrations and Waves,*
> *M.I.T. Introductory Physics Series*

(The first two texts are used at my university.) By the time they reach graduate school, most continuing physics students probably learn that this picture of waves is wrong. However, although it becomes apparent in more advanced texts, the error of this wave picture is seldom pointed out explicitly.* Real waves *can* transport their media. And waves which do are inherently nonlinear, regardless of the medium they occur in.  As we'll see, such waves are actually commonplace in the real world and play a central role in the present theory.

---

*It's been argued that there's nothing wrong with this teaching method. An example cited is the teaching of mechanics, which begins with the inaccurate Newtonian system, superseded later by relativity. There is a difference, though. When relativity is introduced, it's pointed out, repeatedly, that the Newtonian picture is incorrect.

We'll call waves which move their media from place to place "flow waves." Water solitons are a clear example.  The amplitude of an individual "shallow water" soliton propagating in the x direction is given approximately by

$$u = 3\eta sech^2 \frac{\sqrt{\eta}(x-\eta t)}{2} \qquad (1\text{-}1)$$

where u is the water surface elevation in relation to the undisturbed surface, η is the soliton's velocity, and t is time.[6]   This equation indicates that water solitons can hold their shape over time.  For a given t, it describes a symmetrical, raised hump of water, declining to the same level on both sides of the peak, as illustrated in Fig. 1-2.  At the time depicted, most of this soliton is in region A.  Compared to an adjacent region, B, of equal length, A holds more water.  It's apparent that, as the soliton enters B, a water volume equaling that of the hump must flow from region A into B.  Outside the soliton, the water is motionless.  So, as the soliton proceeds further, this displaced water never flows back.

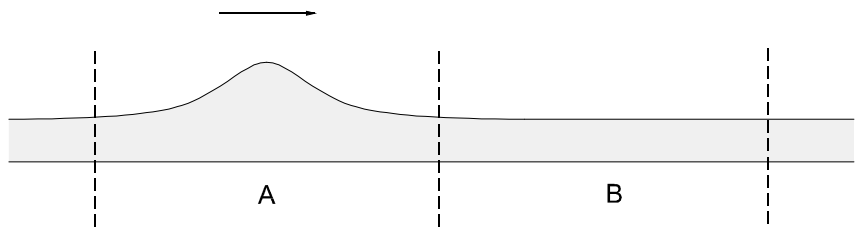

**Fig. 1-2.**  Water soliton.

Naturally occurring solitons can be seen in the ocean photo of Fig. 1-5.  Unlike the isolated soliton just described, this regular array carries an intermittent, but continuing flow of water.  While these apparently were created by a storm, ocean solitons can also be caused by tides.  Oceanographers refer to the overall movement of water brought about by waves as "mass transport."[7]

Solitons are by no means the only flow waves occurring in water.  Ordinary "dispersing" waves can move water from place to place also.  In fact, this is generally the case for the waves created by wind.  Contrary to the first text cited, ordinary ocean waves *do* bring water to the shore.  With large waves, the effects can be dramatic: for example, waves are responsible, to a significant degree, for bringing the high tides associated with storms.

Rip currents are another effect of ocean flow waves.  What happens is this: as waves bring water to the beach and then dissipate, the water level there is raised.  The raised level then drives a return flow back out to sea.  When the incoming waves are small, the return flow usually takes the form of an undertow. However, when the waves are larger, the water draining from the beach often becomes organized into rip currents.[7]

There are many other examples of natural flow waves besides those created by tides or winds. What happens when you toss a rock into a pond? Of course the rock displaces water; most of that displacement is borne off by waves. On a larger scale, tsunamis are flow waves generated by the vertical displacement of the sea floor. When you speak or sing, the sound waves carry away a flow of air. Flow waves also arise in solids. For example, seismic waves propagate lasting displacements in the earth's crust. While they've received scarce attention, waves that transport their media are more the rule in nature than the exception.

Where the present theory is concerned, an important feature of flow waves is that, even in a "linear" medium, they do not superpose like linear waves. When interacting, in addition to displacing each other, flow waves can also refract and reflect one another. We'll be looking at these phenomena later. Except for computer simulation, it is difficult to describe the interactions of flow waves accurately in more than one dimension. However, we'll infer some of their gross behaviors from the interactions of steady flows and waves. (Here we'll be working with acoustic waves. For water waves, an excellent reference on wave/flow interactions is a review by Peregrine[7].)

In addition to its presence in many nonlinear wave phenomena, flow is also found at the heart of a great many chaotic systems. Some well-known examples are Edward Lorenz's air convection model, or the Lorenzian waterwheel (also known as the "turbulent fountain"), dripping water faucets, Taylor vortices, meandering rivers, the Brusselator (a chemical reactor with flowing reactants), piling streams of sand, stirred paint, and Jupiter's Great Red Spot.

## Wraparound Trapping

In 1965, Zabusky and Kruskal[8] published a ground-breaking report on computer simulations of waves in a nonlinear medium. Their paper marked the first use of the word "soliton."\*
The wave system studied was a very simple, one-dimensional one, governed by the

---

\*Although Zabusky and Kruskal were unaware when they introduced the term, solitons were discovered in 1834 by the Scottish scientist and engineer John Scott-Russell. Scott-Russell first observed solitons while studying the waves made by canal barges and later reproduced them in extensive wave tank experiments. Emphasizing the transport, or flow, of the wave medium, his term for the phenomenon was "wave of translation". (Borrowing from Russell, maybe a better term for flow waves would be "translation waves".)

Korteweg-de Vries, or KdV equation:

$$\frac{\partial u}{\partial t} + u\frac{\partial u}{\partial x} + \sigma^2 \frac{\partial^3 u}{\partial x^3} = 0 \qquad (1\text{-}2)$$

This describes the behavior of waves in a dispersive medium, where wave velocity varies both with wavelength and amplitude. The equation pertains (although not precisely) to certain waves in water, plasmas and a variety of other media. (Eq. (1-1) for the amplitude of a water soliton is obtained as a solution of this one.)

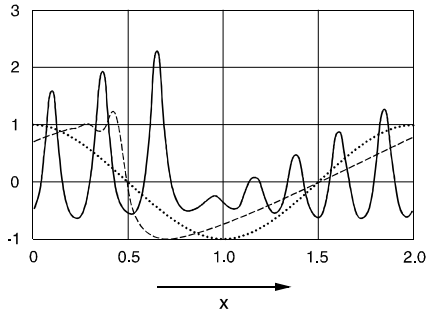

**Fig. 1-3.** Soliton development, from Zabusky and Kruskal.

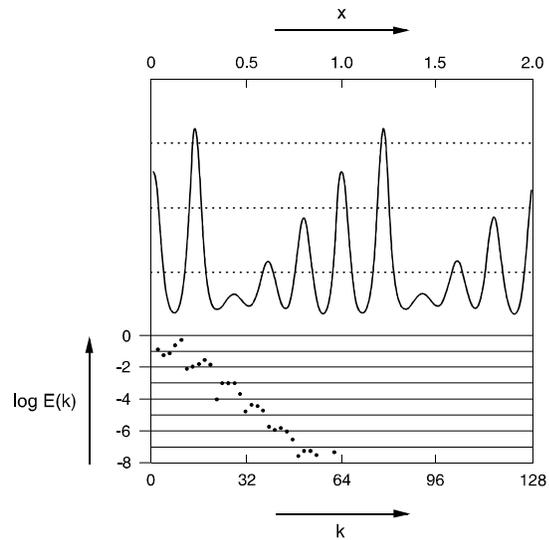

**Fig.** 1-4. A frame of Tappert's soliton movie. (Redrawn.)

Each of these simulations began with a repetitive, sinusoidal waveform, which was subsequently distorted by the nonlinear medium. For small values of σ in the above equation, each cycle of the initial waveform was observed to transform into a number of shorter wave pulses: solitons. Once formed, the solitons persisted indefinitely, "without losing their form or identity," despite interacting and passing through one another. (Previously, it was widely thought that dispersion in nonlinear wave media necessarily leads to the dissolution of localized pulses.) Fig. 1-3 is from Zabusky and Kruskal and depicts a single cycle of a continuously repeating waveform. The dotted line shows the original sinusoidal profile, a solid line the resulting solitons, and the dashed line an intermediate stage.

This wave system was investigated further by Tappert[9]. Like Zabusky and Kruskal, he made a computer-generated "movie" depicting soliton formation. A frame of Tappert's movie, taken from a review by Scott, Chu, and McLaughlin[10] appears in Fig. 1-4. Besides the waves themselves, his movie also shows the energy of each harmonic mode, indicated by its wave number, k. Scott et. al. discuss the figure as follows:

> Observe that for the case of soliton formation, the mode energies fall exponentially with increasing wave number, and consider Planck's formula for the average energy of a mode with quantized energy levels:
>
> $$E(\nu) = \frac{h\nu}{e^{h\nu/k_B T} - 1} + \frac{h\nu}{2} \qquad (1\text{-}3a)$$
>
> $$\approx k_B T, \; for \; \frac{h\nu}{k_B T} \ll 1 \qquad (1\text{-}3b)$$
>
> $$\approx h\nu e^{h\nu/k_B T}, \; for \; \frac{h\nu}{k_B T} \gg 1 \qquad (1\text{-}3c)$$
>
> where $k_B$ is Boltzmann's constant, T is absolute temperature, and h is Planck's constant. The condition for the equipartition of energy is displayed in [1-3b]. The exponential decrease of average mode energy with increasing mode frequency in [1-3c], on the other hand, is qualitatively similar to the effect which forced Planck to assume quantized energy levels. It is intriguing that [Fig. 1-4] displays such an effect *without* the quantum assumption. [Original italics.]

In addition to its seemingly quantum mechanical aspect, the wave system of Zabusky, Kruskal and Tappert is of interest here for another reason: it's related to a multidimensional wave system at the center of the present theory. Both systems are nonlinear and repetitive in one dimension. In systems of this kind, an elementary effect, which we'll call "wraparound trapping," occurs. Referring to Fig. 1-3, if space is divided into regions equal to the period of the initial waveform, you can see that the waves in each region are effectively trapped there. Because of the repetitive nature of the system, every time a wave or wave component exits the right side of a region, an identical wave enters from the left, and vice versa. It's as though each region wraps around on itself. So, the periodicity of such systems is conserved.

**Fig. 1-5.** Water solitons, in a U.S. Army photo from a 1933 *National Geographic*. The original caption read, "As they near the coast of Panama, huge, deep-sea waves, relics of a recent storm, are transformed into waves that have crests, but little or no troughs. A light breeze is blowing diagonally across the larger waves to produce a cross-chop."

**Fig. 1-6.** Waves in an artificial wave tank, showing self-organized structure in the horizontal transverse direction. At the end of the tank where these waves were generated, the wavefronts are straight and smooth.[12] Photo by M.-Y.- Su.

# Coherent Structures

There isn't a standard definition for the term "coherent structure." Most often, it is used in connection with the study of turbulence, where it generally refers to an underlying spatial regularity in the vorticity of turbulent systems. However, it has also been used by nonlinear wave researchers, as a catchall term for any sort of organized nonlinear wave system. In this paper, we'll use a definition which straddles these areas. Here, we define a coherent structure as: a self-organized, regular arrangement of waves and/or vorticity. (We don't require strict periodicity. For example, quasicrystalline wave structures exhibit regularity without being periodic[11].)

A good example of a wave-based coherent structure is the highly regular array of ocean solitons in Fig. 1-5. Rip current patterns are another type of coherent structure. Off long, straight beaches, where the ocean bottom is uniform, rip currents are observed to have fairly regular spacings. Involved are water circulation patterns resembling convection cells. An intriguing aspect of these systems is that the incoming flow waves appear to be diverted by the outgoing rip currents from those regions[7].

Another, very interesting, coherent structure can be seen in Fig. 1-6. The photograph shows a regular arrangement of cusps which develop in artificial waves created in a wave tank[12]. While it normally refers to the vertical dimension of water waves, here we'll use the term "transverse" to indicate the transverse horizontal dimension. In the area of the tank where these waves are generated, no transverse structure at all is seen; initially the wavefronts are straight and smooth. However, as the waves move across the tank, this cusped structure develops. Obviously, without the initial waves, there would be nowhere for cusps to form. With respect to the wave components forming the cusps, *the wave medium consists not of water alone, but of waves as well*.

The wind-blown sand ripples mentioned above are a coherent structure involving both waves and vorticity. In this phenomenon, vortex rolls in the air, caused by the Kelvin-Helmholtz effect, shape the wind and sand. (The rolls form between a moving wind layer and a static boundary air layer at the sand surface.) Concurrently, developing sand ripples promote eddies in the airflow which reinforce the rolls and attract them to the ripples. If the wind doesn't shift, the resulting pattern can be strikingly regular.

An important aspect of coherent structures is that they are *non-local* phenomena, reflecting the overall properties of the physical systems where they evolve. It's apparent, for example, that the shape and size of wind-blown sand ripples are a product of such factors as the general coarseness of the sand, the density of the sand particles, the overall wind speed, etc.

**Fig. 1-7.** Milk drop. Photography by Ronald Bucchino.[13]

It's not always so apparent where the phase of a coherent structure comes from. Starting with a large flat area of sand and a steady wind, sand ripples are observed to form. The question arises: how do they decide *where* to form initially? The answer must be that, while it may be unobservable, at some finer level, there is always some pre-existing structure with a predominant spatial phase. (There is no such thing as a perfectly flat area of sand or a perfectly steady wind.) So the observed large-scale structure and its phase result from the amplification of a lesser, sometimes unseen, structure.

A seemingly magical manifestation of this effect can be found in the high-speed photographs of a milk drop in Fig. 1-7.[13] In the fourth frame, there are eight clearly defined depressions, having a radial symmetry and fairly regular spacings. However, in the first frame, only an instant before, absolutely nothing with an eightfold radial symmetry can be seen. (You could say there are "hidden" variables.) The highly unstable nature of capillary waves (the type involved here) guarantees that such depressions will form. However, it is remarkable how, in no apparent time, the system organizes itself and "decides" the angle at which the overall pattern will be oriented.

This finishes the informal introduction to nonlinear wave phenomena and coherent structures. Where the present theory is concerned, some relevant points are these:

> 1. Flow is an essential element of many self-organizing, chaotic nonlinear systems.
>
> 2. Commonly, waves transport their media, i.e., cause them to flow.
>
> 3. Flow waves are inherently nonlinear, regardless of the medium in which they are found.
>
> 4. Flow waves can refract and reflect each other.
>
> 5. It's not unusual for nonlinear wave systems to organize themselves into repetitive structures.
>
> 6. Repetitive wave structures may exhibit quantum mechanical behaviors.

# 2. A New Wave System

Now I'd like to introduce a new kind of nonlinear wave system. However, first I need to describe the geometric framework in which these waves exist, because it is not the usual Minkowski space-time. Sometimes divergent mathematical representations can be found for the same phenomena. For example, features on the surface of a sphere can be described by either a 2-D non-Euclidean geometry or a 3-D Euclidean one. Here we'll be using a higher-order Euclidean alternative to the non-Euclidean geometry of Einstein and Minkowski.

Two of special relativity's founders, H. A. Lorentz and Henri Poincaré (who was responsible for the term "relativity"), never accepted Einstein's space-time interpretation. Both insisted, to the end, that relativistic phenomena could be accounted for without completely abandoning classical time. To make this possible, Lorentz proposed incorporating relativistic contraction into the theory of matter. However, it was objected that he did so on an ad hoc basis.

Lorentz's theory was also perceived to have a second drawback: It involved a "preferred" frame of reference. Einstein made the compelling argument that it makes no sense to hypothesize a preferred frame if it can't be identified experimentally. We will look at this argument again later. For now, though, I'd like to go ahead and describe a hypothetical physical system which does involve a preferred reference frame, but in which relativistic (as well as quantum mechanical) phenomena arise naturally, without ad hoc hypotheses.

## The Geometric Framework

The framework for this theory is a 4-dimensional Euclidian space and classical time. (So if time is counted as a dimension, this system has five.) We'll call the spatial dimensions x, y, z and $\phi$ (phi). Within this framework there is a repeating system of waves, redundant along the $\phi$ dimension. Initially, we'll assume that the entire wave pattern repeats perfectly along $\phi$ in a sinusoidal fashion, with a spatial period which we'll call the structure interval or s. (In other words, at any given x, y, z location, along $\phi$ we would see a uniform sinusoidal variation in the wave amplitude with a spatial period s.)

Fig. 2-1 shows a simple wavefront pattern in the x,$\phi$ plane. (In a more realistic case, the waves would form interference patterns, without such distinct fronts.) Dividing space into parallel regions of thickness s along the $\phi$ axis, you can see that any wave or wave component exiting one side of a region is matched by an identical wave entering from the opposite side. As in the one-dimensional wave system of Zabusky, Kruskal and Tappert, the waves in each periodic section are effectively trapped there by the wraparound effect, and the periodicity of the overall structure is conserved. (So, if you start with a system of repeating waves with period s, and the waves don't dissipate, any wave patterns which develop will always have this period.)

In describing this 4-D wave system, we'll make frequent use of a 3-D "reference space" corresponding to the locus of points where $\phi = 0$, shown as a horizontal line in Fig. 2-1. (This is analogous to a reference plane in 3-D geometry.)  Because of its repetitive nature, *the entire 4-D wave system can be mapped into this 3-D space*.  In the case of sinusoidal waves, this can be done simply by specifying the sinusoidal amplitude and phase at each 3-D point (x,y,z).  For example, the waveform on the right in Fig. 2-1 depicts the wave amplitude along the vertical line, A, at point $(x_A, y_A, z_A)$.  The waveform can be mapped to this point in the reference space by its overall sinusoidal amplitude, plus a phase value determined by the waveform's positioning along $\phi$.  (The symbol $\phi$ was chosen because it often represents phase.)  Mapping this four-dimensional system into three will let us compare it with the three-dimensional world we experience.

**Fig. 2-1.** A wavefront pattern and associated amplitude at $x_A$, $y_A$, $z_A$ in the reference space.  Vertical line A represents a "pro-particle."  (While the wavefronts are hypersurfaces, A is only a line in four dimensions.)

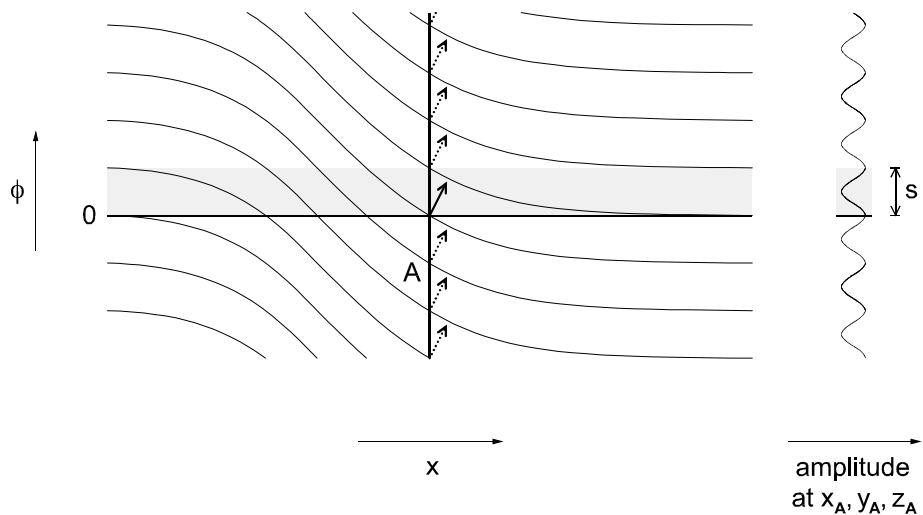

## The Wave Medium

The wave medium in this theory is an elastic fluid; consequently, its waves are acoustic (i.e. sound-like).  Except for its four-dimensional aspect, in many other respects it resembles an ordinary fluid, such as air.  For example, a box of air is inhomogeneous and chaotic at the scale of molecules, but relatively uniform at the macroscopic level.  Similarly, the properties of this wave medium will be taken to vary at different scales.  In its scale-dependant nature, this medium differs from the historical ether, which was generally considered "ideal."*

---

*In Greek mythology, the ether was something pure in the rarified atmosphere of Mount Olympus, breathed by the gods.

There are several other differences:  In theories such as Lorentz's "Theory of the Electron," the ether was assumed to be an elastic solid, to accommodate transverse electromagnetic waves.  Of course it had one less spatial dimension.  Also, the ether was generally taken to be distinct from "ponderable matter," and extemely tenuous, so the latter could pass through it unimpeded.  On the other hand, the wave medium in this theory is substantive.  Further, elementary particles here are wave-based phenomena, where both they and their medium are constituted of essentially the same stuff.  So I'd like to call this four-dimensional fluid wave medium simply "stuff."

Like sound in air at a given temperature, the acoustic waves in stuff have a characteristic velocity -- in this case, the speed of light, c.   Since such waves are essentially longitudinal, of course they can't represent light per se, although they travel at more or less the same speed.  (How can electromagnetic waves be accounted for then?  As we'll see later, in repetitive nonlinear wave structures, transverse secondary waves can arise.  These are waves of interaction *within* the waves, or *wave waves* if you like!  Already we've partially glimpsed such behavior in the cusped water waves of Fig. 1-6.)

In this "4-D Wave Theory," acoustic waves are responsible not only for particles, but also their fields.  Before describing how, I need to introduce various basic features of the wave system.  However, to introduce this system, it seems that something resembling particles is needed.  So, initially, I'd like to make use of a very incomplete particle model proposed by Schrödinger.

## Pro-particles and de Broglie Waves

In the early 1800's W. R. Hamilton developed a new formulation of mechanics based on a close analogy between the behavior of material bodies and light.  His system was based on hypothetical "surfaces of constant action," analogous to light wavefronts, whose normals (rays) described the motions of material bodies.  Schrödinger was inspired by Hamilton's optical-mechanical analogy in his development of wave mechanics, and felt that it might be extended to further account for the elementary particles.  It was Schrödinger's contention that particles might be nothing more than geometric normal rays associated with the continuous wavefronts of matter waves[14].  In the present theory, elementary particles are not such simple entities.  However, at this point it will be useful to assume Schrödinger's ray/particle model and explore its implications in the context of  this wave system.  (Many features of this simplified particle model will carry over.)

One of these Schrödinger "particles" can be seen as a point moving with a wave, along a normal to the wavefront, keeping the same phase position on the wave.  (As though it were somehow possible to mark a point on a wave itself, instead of marking the wave medium.)  The trajectory or velocity vector of this wave-point corresponds then to a ray.

In Fig. 2-1, the bold arrow at $x_A,0$ depicts the 4-D velocity vector of a Schrödinger "particle" in our new wave system. Notice that, from the repetitive nature of this system, we necessarily have identical "particles" and vectors at $\phi = s$, $\phi = 2s$, etc., shown here as dotted arrows. Also, between these are others, which, although differing in phase, have exactly the same velocity vector. Since all these "particles," a continuous line of them at a given x, y, z position, move together, we can treat them as a single unit. This is depicted by the line, A, in the figure, which parallels the $\phi$ axis.

We'll call these vertical, line-shaped ensembles "pro-particles," short for "provisional particles." (We'll be working with pro-particles until we get to the realistic particle model introduced in a later section.) While it corresponds to a line in four dimensions, a pro-particle appears as a single point in our 3-D reference space (having an associated phase value). Note that a pro-particle's position in the reference space is unchanged by its motion in the $\phi$ dimension. So, with waves moving parallel to the $\phi$ axis, the observed velocity of an associated pro-particle is zero. On the other hand, if the waves are moving almost perpendicularly to $\phi$, the pro-particle velocity seen in the reference space approaches its 4-D velocity, c.

**Fig. 2-2.** Two sets of hyperplane waves, with different wavelengths, as seen in the reference space ($\phi = 0$).

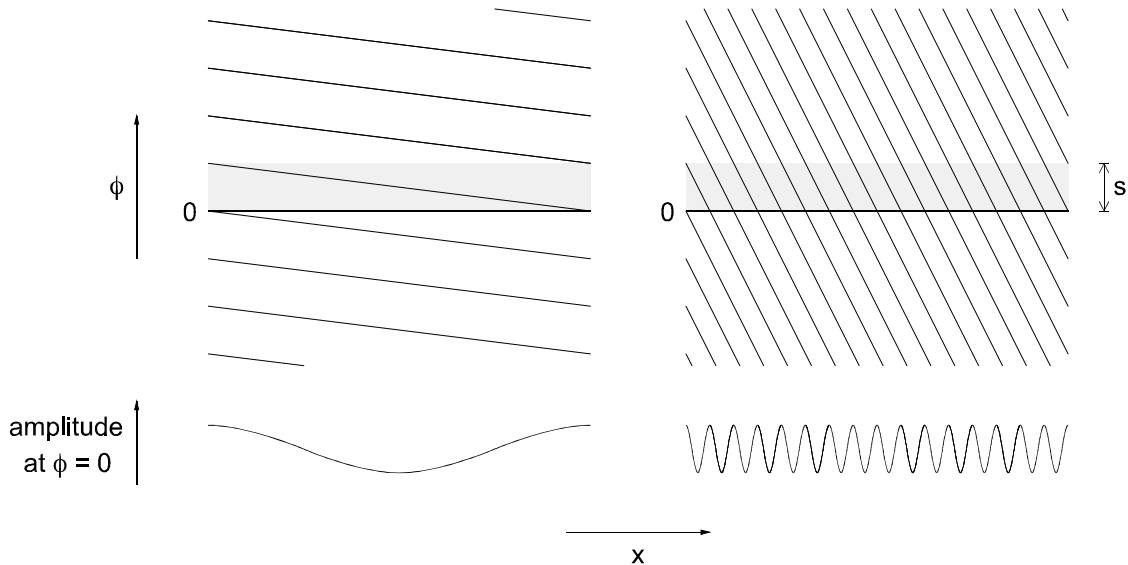

Our wave system allows the four-dimensional equivalent of 3-D plane waves; we'll call these hyperplane waves. (We can refer to such a geometric entity as a 3-D "space," like our "reference space," or a 4-D hyperplane, when we want to emphasize the analogy to a 3-D plane.) Two sets of hyperplane waves are depicted in Fig. 2-2. Above are the wavefronts in the x,$\phi$ plane and below are the corresponding wave amplitudes seen in the reference space, $\phi = 0$. Although the wavelengths are different, both sets of waves necessarily have the same spatial period, s, along the φ dimension.  Arbitrary waveforms, including the curving wavefronts depicted in Fig. 2-1, can be constructed from multidimensional Fourier series' of sinusoidal hyperplane wave components such as these.

The relativistic wavelength of the de Broglie waves associated with an actual particle of rest mass $m_0$ is given by

$$\lambda = \frac{h\sqrt{1-v^2/c^2}}{m_0 v} \tag{2-1}$$

where v is its velocity, and h is Planck's constant.  (With no external vector potential.)  What wavelength is associated with a moving pro-particle?  Here we'll refer to the two-part diagram of Fig. 2-3.  On the left side, we have a set of hyperplane wavefronts, moving up and to the right.  The vertical line, A, represents a pro-particle associated with the waves.  Again, s is the φ interval for the waves, while λ is the wavelength seen in the reference space.

**Fig. 2-3.**  This shows the relationship between the velocity, v of a pro-particle and the wavelength, λ, of its associated waves.  The vector with magnitude c represents the 4-D pro-particle velocity and is perpendicular to the wavefronts, while v is its component in the x direction.

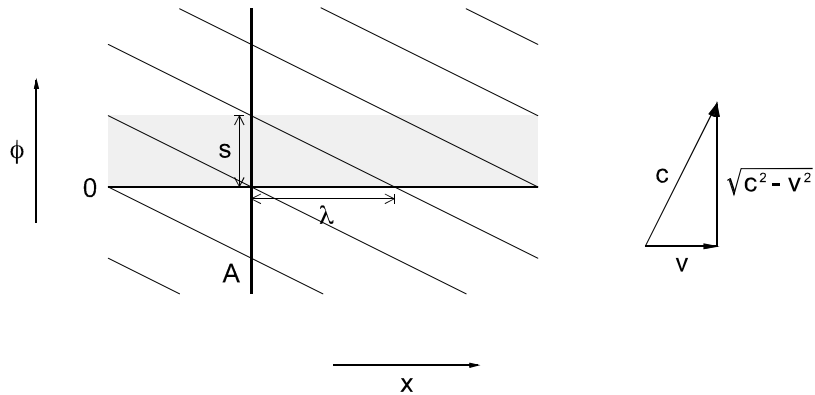

The vector triangle on the right corresponds to the pro-particle's motion.  In diagrams like this one, unless in bold-faced, the vector labels will refer to their magnitudes (rather than identifying the actual vectors themselves).  The vector with magnitude c represents the 4-D pro-particle velocity and is perpendicular to the wavefronts.  The vector labeled v is the component seen in the reference space.  Similar triangles give

$$\frac{\lambda}{s} = \frac{\sqrt{c^2-v^2}}{v} \tag{2-2}$$

and

$$\lambda = \frac{sc\sqrt{1-v^2/c^2}}{v} \qquad (2\text{-}3)$$

Notice that, if we take s equal to the constant $h/m_0c$, this gives Eq. (2-1), the exact wavelength for a moving particle of mass $m_0$.

The phase velocity, V, of the de Broglie waves associated with an actual particle is given by

$$V = c^2/v \qquad (2\text{-}4)$$

What phase velocity is associated with a pro-particle? Next we'll refer to Fig. 2-4, which shows the same waves, pro-particle, and vector triangle as Fig. 2-3. Here, vectors with magnitudes c and V are added on the left. Again, c is the 4-D wavefront velocity, while V is the phase velocity of the wavefronts seen in the reference space.

**Fig. 2-4.** This shows the relationship between pro-particle velocity, v, and the phase velocity, V, of associated waves. Both vectors, c, are perpendicular to the wavefronts.

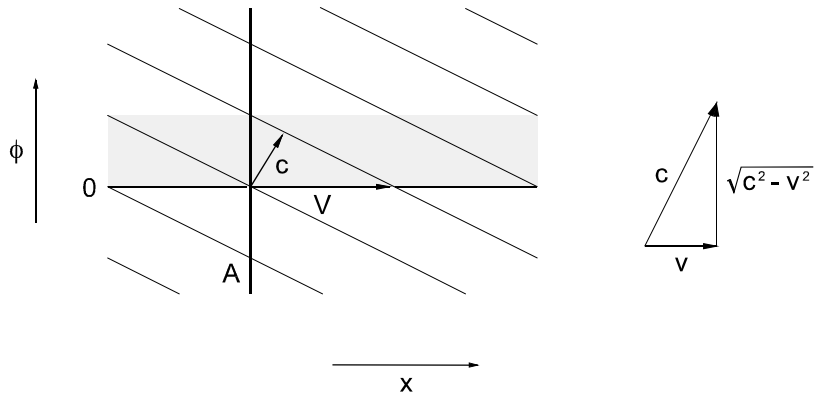

From similar triangles, we have

$$\frac{V}{c} = \frac{c}{V} \qquad (2\text{-}5)$$

or

$$V = c^2/v \qquad (2\text{-}6)$$

So this model gives the correct phase velocity for an actual particle of any mass.

Eq. (2-4) says that de Broglie waves have a phase velocity which is at least c, and is infinite for motionless particles. In the accepted picture, these phase velocities are considered the true speeds of these waves, meaning that they are fundamentally different from ordinary ones. However, *in this 4-D system, instead of infinite wave velocities, we simply have spatially extended wavefronts*. Like water waves intersecting a beach, there is no limit to how fast a phase point can move along, as different parts of a wavefront cross a line of reference. We'll see that, although four-dimensional, the waves in this theory are much more closely related to familiar ones than those of present quantum mechanics.

An equation we'll be using later is that for the wave frequency associated with a pro-particle. Dividing the phase velocity V by the wavelength $\lambda$, we have

$$\nu = \frac{c}{s\sqrt{1-v^2/c^2}} \qquad (2\text{-}7)$$

where $\nu$ is the frequency seen at a fixed position in the reference space. This increases with v.

Also of interest is the frequency seen at the moving position of a pro-particle. In this case, the frequency of these waves corresponds to the velocity of a pro-particle in the $\phi$ direction, divided by the $\phi$ period, s. Referring to the vector triangle on the right side of Fig. 2-4, this velocity component is $\sqrt{c^2-v^2}$. Calling the frequency at the pro-particle $\nu_v$, we have:

$$\nu_v = \frac{\sqrt{c^2-v^2}}{s} = \frac{c}{s}\sqrt{1-v^2/c^2} \qquad (2\text{-}8)$$

So, the greater the pro-particle velocity, the *lower* the wave frequency seen at its moving position. For a stationary pro-particle, where the waves move parallel to the $\phi$ axis, the value is c/s. Calling this $\nu_0$, the frequency seen at a moving pro-particle is then

$$\nu_v = \nu_0 \sqrt{1-v^2/c^2} \qquad (2\text{-}9)$$

As it should, this corresponds to the $1/\sqrt{1-v^2/c^2}$ time dilation factor given by relativity.

# 3. Special Relativity

Although the physical system in this theory is properly relativistic, a preferred frame of reference is involved. So, this section begins with an overview of the preferred frame issue.

In solving relativity problems, one typically chooses some particular reference frame with a given state of motion, to which events in other frames are referred. Observed from this "rest" frame, clocks in all others run more slowly, and objects in all other frames are contracted. Also, as seen from this chosen frame, the various relativistic phenomena occurring in other frames interact such that these effects are not detectable by observers in those frames.

Lorentz and Einstein came to fundamentally different interpretations as to the physical reality behind this situation. To preserve the classical role of time in the universe at large, Lorentz argued that there is some reference frame, corresponding to a state of absolute rest, in which time is not dilated. Viewed from this preferred frame, events in other frames are seen to behave exactly as they do when one arbitrarily chooses a rest frame in solving a relativity problem. However, the relativistic effects observed from this frame are considered objectively real. As Lorentz repeatedly pointed out, such an approach is equivalent to Einstein's space-time interpretation from the standpoints of both mathematics and experiment. (At least as far as special relativity is concerned.)

Einstein argued that since all reference frames are subjectively equivalent (with respect to linear motion), it necessarily complicates things to assume that one is preferred. Much of the force of this argument comes from the idea that a preferred frame system requires the ad hoc assumption of relativistic transformations. However, Einstein's space-time interpretation *does* include a preferred frame for rotational motion. (While linear motion is defined by equivalent reference frames derived from space-time geometry, Einstein postulated that rotational motion is defined by "the fixed stars" -- a preferred frame.) So, while preferred frames are not eliminated, there is the additional complication of different frames for linear and rotational motion. Another difficulty with the space-time interpretation of relativity is that it does not account for the direction of entropy; an additional time-like factor of some sort seems to be required.

In this theory, as in Lorentz's "Theory of Electrons," one consistent, preferred frame of reference defines both linear and rotational motion. Also, both incorporate classical time, which can be taken as directional in nature, giving a preferred direction for entropy. However, in this theory, all relativistic transformations spring directly from the physical system involved, and need not be assumed ad hoc, as Lorentz did for relativistic contraction.

For a reference frame moving in the x direction, the Lorentz transform equations are:

$$x' = \frac{x - vt}{\sqrt{1 - v^2/c^2}} \tag{3-1}$$

$$y' = y \tag{3-2}$$

$$z' = z \tag{3-3}$$

$$t' = \frac{t - vx/c^2}{\sqrt{1 - v^2/c^2}} \tag{3-4}$$

Of course, within the context of a preferred frame system, Eq. (3-1) means that objects actually contract along their direction of motion. Eq. (3-4) means that moving clocks run more slowly in an absolute sense. And, moving clocks which are separated along their direction of motion are desynchronized by an absolute amount of time.

For those unaccustomed to preferred frames, I'd like to illustrate their nature with a physical analogy. It's a (tongue-in-cheek) thought experiment involving sound: Three nearsighted cowboys are riding atop a train, in the open air. Our men are on different cars, with the length of five identical cars separating each of them. The two outside cowboys want to synchronize their watches, but their distance vision is much too poor to use visual signals.

So they call to the middle cowboy to shoot his gun, and both start their watches when they hear the shot. However, because the train is moving, the sound must travel farther through the air to reach the forward cowboy than it does to reach the man in back. The cowboy in the middle, a bright fellow named Al, realizes there may be a problem. So, to check their synchronization, he has the other cowboys shoot their guns when their watches read exactly one o'clock. What does he hear?

Al hears two simultaneous shots. On that basis, he decides that the watches really are synchronized, and also concludes that the speed of sound outside is invariant with respect to moving trains. His mistake, of course, was forgetting that the sound from the rear had to travel farther through the air to reach him than the sound from up ahead. This exactly cancels the discrepancy of the two watches. So, he really has *no way of knowing*, from the timing of sounds, whether the watches are actually synchronized or not.* This is precisely the way light behaves in preferred reference frame systems, such as the ether-based model of Lorentz.

---

*Some wonder whether Al might hear a difference in pitch between the two sounds, due to Doppler shifting. Of course he would if he were stationary with respect to the air. However, because of his motion, Al experiences Doppler shifts at the receiving end which cancel those at the sources. For example, the sound from ahead is shifted to a lower absolute pitch, but he intercepts these waves at an increased rate.

It was Poincaré who first noted that, in a system like Lorentz's, if an observer's senses were mediated entirely by electromagnetic waves, he would have no way of discerning relativistic effects within his own reference frame, and no way of knowing how fast his frame is moving relative to the ether. Lorentz's theory does contain the ad-hoc assumption that moving particles and objects contract along their direction of motion. However, given this, the remaining effects of special relativity result naturally from the behavior of electromagnetic waves traveling at a uniform speed through a fixed ether.

In its treatment of light, Lorentz's theory is an elegant one. Although the contraction of material objects is assumed, this assumption doesn't pertain to light; the lengths of light waves are naturally correct as the result of Doppler shifting. Time dilation automatically results from these waves traveling increased distances through the ether in moving frames. And, time desynchronization in moving frames results from light waves traveling with differing relative velocities in different directions, just like the sound waves in our cowboy analogy.

(For a more complete discussion of Lorentz/Poincaré relativity, see J. S. Bell[15]. Bell argues that their preferred-frame approach is preferable for teaching special relativity, and reiterates its experimental equivalence to Einstein's space-time. He also describes how, for a classical atom with a non-radiating electron, Lorentz's contraction assumption can be avoided by invoking Maxwell's equations.) Having noted some appealing features of preferred frames, now we'll take up relativity in the context of our new wave system.

As shown earlier, when this 4-D system is viewed from the perspective of a 3-D reference space, the waves associated with pro-particles resemble de Broglie waves. De Broglie based the equations for his waves on Einstein's relativity; consequently, they are inherently relativistic with respect to both space and time. The waves in this new system will be seen to behave in the same relativistic fashion. Also, we'll find that this automatically brings relativistic behavior to our provisional particles (as well as to the realistic particles introduced later).

Recall that a pro-particle corresponds to a line in Euclidean 4-space, paralleling the $\phi$ axis, and represents points of varying phase on redundant waves. (Or a single point in our 3-D reference space.) The fronts of the associated waves all make the same angle with respect to the line of the pro-particle. And the points making up the 4-D pro-particle line all have the same velocity vector, so they move together. As we've seen, for a pro-particle to move in the 3-D reference space, its associated wavefronts must be tilted with respect to $\phi$. Here we'll look at some hyperplane wavefronts and assume that they are tilted by some process which leaves the length of a given section of wavefront unchanged. (A mechanism which tilts waves in this fashion will be described later.) What effect would such tilting have on an arrangement of our pro-particles?

Figs. 3-1(a) and (b) show a pair of pro-particles, A and B, associated with the same hyperplane waves. (As before, vectors on the right represent the pro-particle motions.) In (a) the wavefronts are perpendicular to $\phi$, and A and B are stationary in three dimensions. Fig. 3-1(b) shows the same wavefronts, tilted, with the pro-particles moving at velocity v. The length of wavefront separating A and B in four dimensions (invariant by our assumption) is $d_0$, while d is the apparent distance seen in the reference space, $\phi=0$.

**Fig. 3-1.** The same waves, tilted and untilted, are shown. Pro-particles A and B, associated with the waves, are separated by the same wavefront length, $d_0$, in each case. Distance d separating A and B in the reference space contracts with tilting.

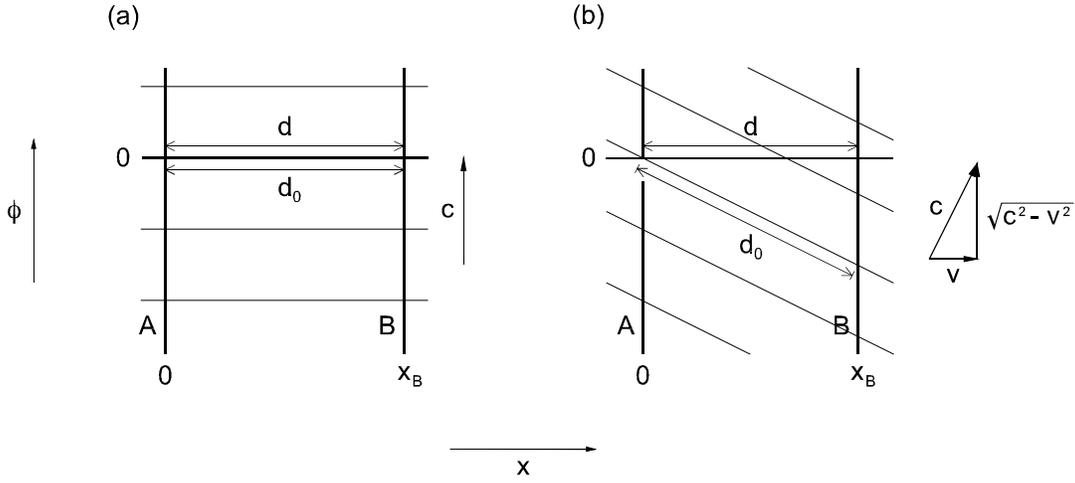

Referring to Fig. 3-1(b), by similar triangles, we have:

$$\frac{d}{d_0} = \frac{\sqrt{c^2 - v^2}}{c} \qquad (3\text{-}5)$$

and

$$d = d_0\sqrt{1 - v^2/c^2} \qquad (3\text{-}6)$$

The factor $\sqrt{1 - v^2/c^2}$ corresponds to Lorentz-Fitzgerald contraction. As you might expect, Eq. (3-1) can be derived from this effect. To do this, we'll put things in terms of the usual two coordinate systems. Unprimed coordinates x, y, z, and t will refer to our preferred frame and coordinates x', y', z', t' to a frame moving with the pro-particles. We let the x and x' origins correspond to the position of A at time t = 0. In our preferred frame, the x coordinate of pro-particle B is then given by

$$x = d_0\sqrt{1 - v^2/c^2} + vt \qquad (3\text{-}7)$$

When v = 0, we have d = $d_0$ = x = x'. If measurements are made in terms of pro-particle spacings, then within a moving reference frame, there are no measurable differences in lengths or distances at different frame velocities. (This is because any arrangement of pro-particles that might be used as a measuring stick contracts along with any other.) So, we can take x' to be invariant. If x' and $d_0$ are both invariant, and equal when v = 0, we also have x' = $d_0$. Substituting for $d_0$ in Eq. (3-7) gives

$$x = x'\sqrt{1 - v^2/c^2} + vt \qquad (3\text{-}8)$$

and

$$x' = \frac{x - vt}{\sqrt{1 - v^2/c^2}}$$

the first Lorentz transform equation.

If a hyperplane is tilted in the x and φ dimensions, there is no change in the y and z coordinates of points on the hyperplane. So, for pro-particles moving only in x, where the wavefronts tilt only in x and φ, y and z are unaffected by the velocity. This gives us y' = y and z' = z, the second and third Lorentz transform equations.

Earlier, we found that, at the moving position of a pro-particle, the frequency of its associated waves decreases as $\sqrt{1 - v^2/c^2}$, corresponding to relativistic time dilation. If we assume that a pro-particle effectively measures time by the phase of its waves, we can derive Eq. (3-4), the fourth Lorentz transform equation. (The assumption will be borne out later.)

**Fig. 3-2.** Untilted and tilted waves again. The waves on the left have the same phase in the reference space at A and B. Those on the right lag in phase at B.

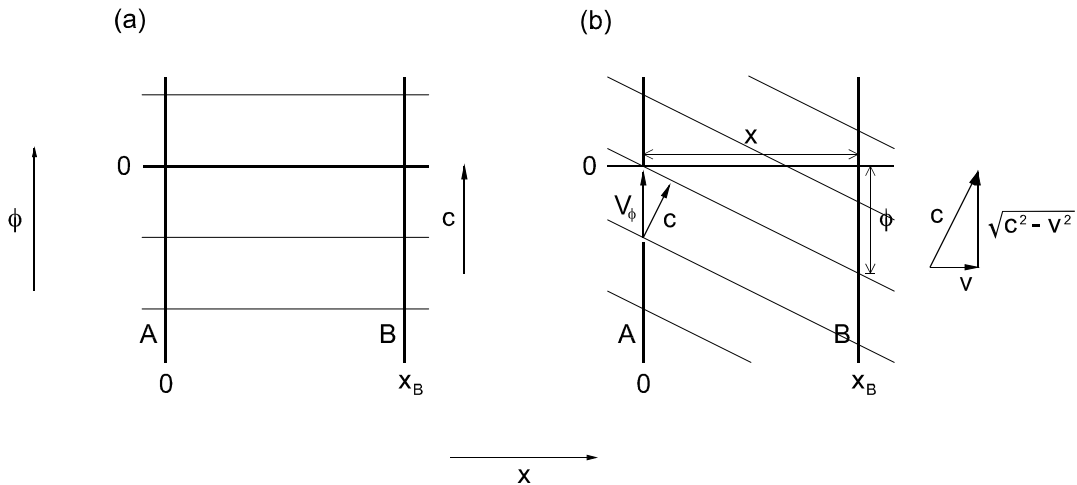

Fig. 3-2 shows the waves and pro-particles of Fig. 3-1 once again. As depicted on the left, when the pro-particles are stationary in the reference space, the waves have the same phases at A and B. With moving pro-particles, shown on the right, a wavefront crossing the reference space at A won't reach it at B until some later time. So pro-particle B lags A in phase.

Here we'll write an equation describing a representative wavefront, in preferred frame coordinates, as a function of pro-particle velocity. At time t = 0, we let the wavefront cross the reference space at x = 0 (where we've placed pro-particle A). From the vector triangle on the right in Fig. 3-2(b), the slope of the wavefront is given by $-v/\sqrt{c^2 - v^2}$. At t = 0, the equation of the wavefront is then

$$\phi = \frac{-vx}{\sqrt{c^2 - v^2}} \tag{3-9}$$

As this wavefront travels, its φ intercept (its intersection with a vertical line at x = 0) moves upward. We'll call this intercept movement the φ phase velocity, or $V_\phi$. A vector triangle corresponding to $V_\phi$, and the wavefront velocity, c, has been added on the left side of Fig. 3-2(b). From this and the vector triangle on the right, it can be seen that

$$\frac{V_\phi}{c} = \frac{c}{\sqrt{c^2 - v^2}} \tag{3-10}$$

and

$$V_\phi = \frac{c^2}{\sqrt{c^2 - v^2}} \tag{3-11}$$

Eq. (3-9) is the equation of a line with a φ intercept of 0 at time t = 0. From this and Eq. (3-11), which gives the change in intercept, we can write an equation for the wavefront at an arbitrary time t:

$$\phi = \frac{-vx}{\sqrt{c^2 - v^2}} + V_\phi t = \frac{-vx + c^2 t}{\sqrt{c^2 - v^2}} \tag{3-12}$$

In the case where the velocity v is zero, depicted in Fig. 3-2(a), this reduces to

$$\phi = ct \tag{3-13}$$

When the velocity is zero, of course the preferred and "moving" reference frames are the same, and in primed (moving frame) coordinates we also have

$$\phi = ct' \tag{3-14}$$

In terms of pro-particles themselves, there can't be any measurable time dilation within a moving frame. (Since any pro-particles or waves that might be used as a clock slow down along with everything else.) So, this last equation can be taken to hold for any velocity. Substituting this expression for φ in Eq. (3-12) gives

$$ct' = \frac{-vx + c^2 t}{\sqrt{c^2 - v^2}} \tag{3-15}$$

Which can be rearranged as

$$t' = \frac{t - vx/c^2}{\sqrt{1 - v^2/c^2}}$$

the fourth Lorentz transform equation.

# Moving Mirrors

So far, no mechanism has been suggested for the wave transformations we've been exploring. Here, we'll take this mechanism to be reflection. In deriving the Lorentz transform above, two basic assumptions were made. The first was that our waves are tilted by some means which leaves the length of a hyperplane wavefront section unchanged. The second was that, somehow, pro-particles effectively measure time by the phases of their associated waves. Below, we'll see that these assumptions can be superseded by a single, more definitive one: that all our waves have been reflected at some point.

To obtain our wave transformation, our basic approach will be to ask what would happen to these waves if the "mirrors" from which they reflected had had different velocities. We'll see that, when the mirror velocity is changed, the reflected waves are transformed such that, within another relativistic frame, they have exactly the same configurations and states of motion that they otherwise would have had. In other words, the effect of this "moving mirror transform" is a complete Lorentz transformation.

What sort of "mirror" are we talking about? In keeping with the rest of this theory, the mirrors we will be considering here are themselves wave-based phenomena. While the physical reflection mechanism won't be addressed until a later chapter, the point here is this: Unlike "solid" objects, these wave-based mirrors can pass through the wave medium without having to push it out of the way. For example, suppose you had a sound-reflecting concrete wall moving through air; the induced airflow *around* the wall would influence the propagation of reflecting sound waves. Here, we won't be concerned with such an effect.

Since our wave system is redundant along $\phi$, any wave-based mirror surface that may arise must have an overall orientation parallel to this axis. (That is, on a scale larger than the wave structure's period, s.) We'll also add the working assumption that our mirror surfaces are locally straight and continuous along $\phi$. (This will be justified later, when the physical reflection mechanism is discussed.)

To begin, we need a reflection law for our moving mirrors. For a stationary mirror, there is the familiar law of reflection: the angle of reflection equals that of incidence, with angles measured with respect to a surface normal. This no longer holds when the mirror moves. Here we'll use Huygens' principle to find a more general law, while making some qualitative observations.

Fig. 3-3(a) depicts a hyperplane wavefront being reflected by a moving mirror. The mirror is a also hyperplane, perpendicular to the x axis, moving in the positive x direction with velocity w. A broken vertical line, $M_0$, represents the mirror at time t = 0, while the solid line, $M_1$, shows the mirror at a later time, t = 1. Both the incident and reflected waves are perpendicular to the x,$\phi$ plane. The bent angled lines, $F_0$ and $F_1$, represent the wavefront at times t = 0 and t = 1 respectively. The upper sections of $F_0$ and $F_1$ are moving up and to the left, while the lower sections, having been reflected, are moving up and to the right.

**Fig. 3-3.** Reflection by a moving mirror. In (a), $M_0$ and $M_1$ depict a mirror at times t=0 and t=1, respectively. $F_0$ and $F_1$ are a reflecting wavefront at these times. The corresponding points of contact between wavefront and mirror are $P_0$ and $P_1$. In (b), the mirror is omitted and the mirror/wavefront contact surface, S, is shown. Also added are normals to S at $P_0$ and $P_1$. A is an incident and B a reflected ray.

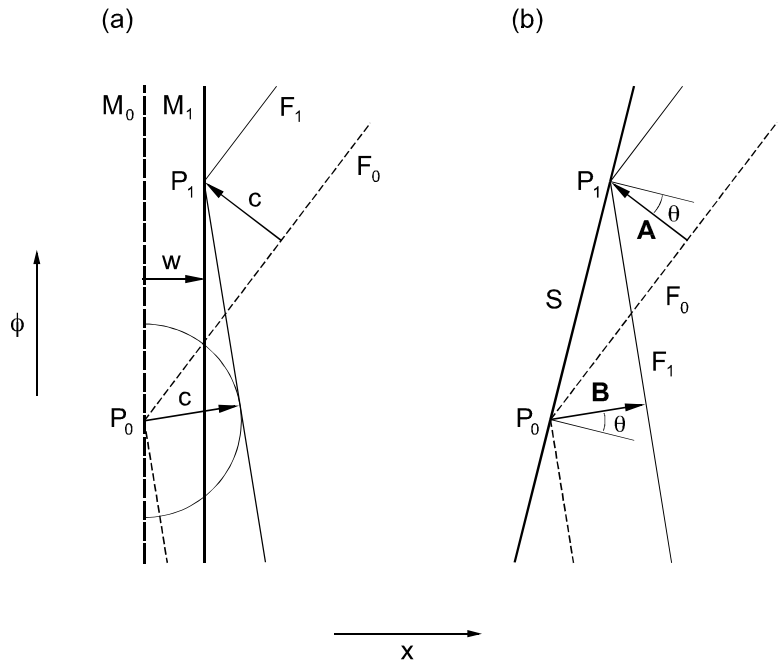

**Fig. 3-4.** Another case of reflection by a moving mirror. Instead of moving toward it, here the mirror retreats from the incident wavefront.

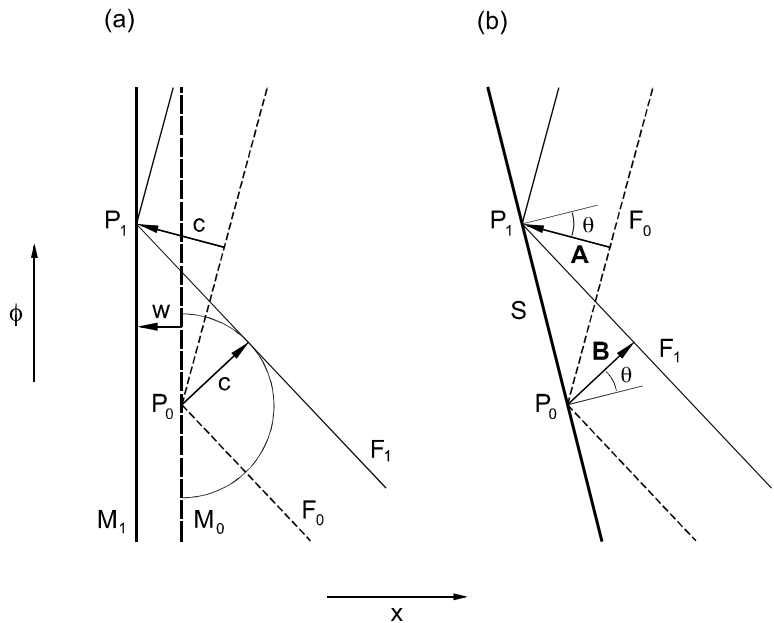

As shown in Fig. 3-3(a), the wavefront contacts the mirror at $P_0$ at time t = 0, and at $P_1$ at t = 1. A Huygens' wavelet originating from $P_0$ at t = 0 becomes the semicircle shown at time t = 1. By Huygens' principle, the lower section of wavefront $F_1$ must be tangent to this wavelet and pass through $P_1$. A vector with magnitude c depicts the wavelet expansion and also the wavefront speed. A smaller vector, with magnitude w, depicts the relative mirror velocity. You can see that, for an increasing w, point $P_1$ is shifted to the right of $P_0$. This causes the reflected wavefront to approach a vertical orientation and reflected rays to approach the horizontal, closer to a mirror normal (not shown).

A second case is illustrated in Fig. 3-4(a). Unlike the last, where the mirror moves toward the incident wavefront, here it retreats, in the -x direction. As a result, $P_1$ is shifted to the left of $P_0$. Here, the reflected wavefront has a less vertical orientation, and the reflected rays are less horizontal, deflected away from a mirror normal. Summarizing, *when a mirror moves toward incident waves, the reflected rays are brought closer to a mirror normal. Or, when the mirror retreats, reflected rays are shifted away from the normal.*

Figs. 3-3(b) and 3-4(b) show the same mirror/wavefront interactions depicted in parts (a). In parts (b), the mirror surface is not shown and the wavefront motion vectors are labeled **A** and **B** instead of c. **A** and **B** can be taken to represent the movements of a pair of pro-particles associated with the wavefront. (Of course these vectors are normal to the wavefront.) The tails of these vectors lie on $F_0$ and correspond to the positions of these pro-particles at time t = 0, while the heads lie on $F_1$ and correspond to t = 1. (Following previous diagrams, the pro-particle positions at times t = 0 and 1 might have been illustrated with vertical lines. These are omitted to keep the figure readable.)

Added in Figs. 3-3(b) and 3-4(b) is a line, S, running through points $P_0$ and $P_1$. S represents the surface where, at different times, the mirror and wavefront make contact. (S differs for different wavefronts.) Also added are a pair of normals to S, at points at points $P_0$ and $P_1$. From the identical triangles bounded by $F_0$, **A**, S and $F_1$, **B**, S, you can see that the angle θ between a normal to S and the incident pro-particle vector, **A**, equals the angle between a normal and the reflected pro-particle vector, **B**. *What we have, then, is just the ordinary law of reflection, except that, for each wavefront, we take as our mirror the mirror/wavefront contact surface.*\* This is the generalized law for moving mirrors. It says that what matters is not the mirror orientation per se, but where each point of a mirror is when a wavelet reflects from it.

Now we're ready to derive our wave transformation. For simplicity, we'll be working with hyperplane mirrors. (The same wave transformation also occurs with other mirror shapes.) We'll choose our coordinate system such that the x dimension is perpendicular to the mirror surface. In whatever direction the actual mirror may move, the velocity of its surface then can then be given entirely in terms of its x component. (This is because any y, z, or φ velocity components then are parallel to the surface and don't change its effective positioning.)

---

\*Of course there is a second difference between ordinary reflection and reflection from moving mirrors: Doppler shifting. In addition to being tilted, wave trains are compressed and expanded, exactly like the wavefronts shown in Fig. 2-2.

Given that the mirror is a hyperplane, we can immediately address one of our earlier assumptions: that our waves are tilted by some means which leaves the length of a hyperplane wavefront section unchanged. When the moving mirror and wavefronts are both hyperplanes, the mirror/wavefront contact surface is a hyperplane also. In this case, the effective mirror is flat, of course, and reflection by a flat mirror doesn't change the length of a wavefront section.

Next, we'll begin to address the other assumption made in deriving the Lorentz transform - that pro-particles effectively measure time by the phases of their waves. In theories with preferred reference frames, like this one, time itself is unchanged in moving frames. What changes instead is the *behavior* of things *in time* -- velocities in particular. (For example, the hands of a moving watch turn more slowly, in an absolute sense, than those of a watch at rest.) Below, we'll show that pro-particle velocities are transformed relativistically by the motion of a reflecting hyperplane mirror. Since their velocities necessarily determine the rates by which systems of pro-particles would gauge time, the previous assumption concerning wave phases will be unnecessary.

First, from our generalized law of reflection, we'll derive a set of equations for the velocities of pro-particles reflected by moving mirrors. Unlike the two-dimensional work done so far, here we'll describe all four dimensions of our system this. (Consequently, the derivation is a little long.) Then we'll use these equations to give us those for the relativistic transformation of velocity.

# Moving Mirrors in 4-D

Fig. 3-5(a) shows the basic 4-D vectors we'll use to write a set of velocity equations for reflected pro-particles. Each vector corresponds to a velocity with magnitude c.* **A** represents the motion of an incident pro-particle, while **C** is the same pro-particle after reflection. Also shown is a slanted line, S, which again represents the mirror/wavefront contact surface. Vector **B** is an artificial construct, normal to this fixed surface. (**B** doesn't correspond to the movement of anything.) As with ordinary reflection, from our generalized law, **-A**, **B**, and **C** are separated by equal angles and lie in a plane.

**Fig. 3-5.** Vectors used to describe 4-D reflection. S is the mirror/wavefront contact surface. On the right, vector **C** is displaced to show its relationship to **A** and **B**.

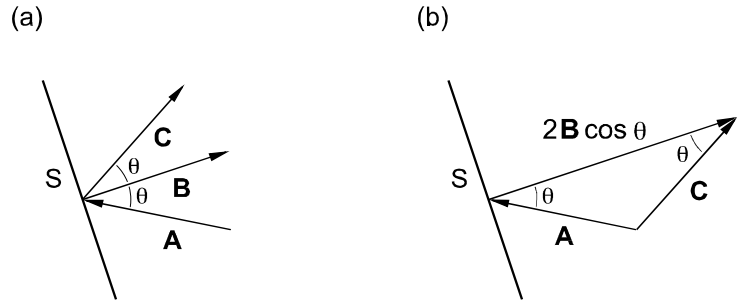

Given **A** and the mirror movement, we'll construct vector **B**. Then, as illustrated in Fig. 3-5(b), we can use the equation

$$\mathbf{C} = \mathbf{A} + 2\mathbf{B}\cos\theta \tag{3-17}$$

to find **C** from vectors **A** and **B**.

The plane of **A**, **B**, and **C** is generally skewed with respect to our coordinate axes, so these vectors have components in all four dimensions. The components of **A** are

$$\mathbf{A} = \mathbf{i}A_x + \mathbf{j}A_y + \mathbf{k}A_z + \mathbf{l}A_\phi \tag{3-18}$$

where **i**, **j**, **k**, **l** are unit vectors in the positive x, y, z, $\phi$ directions.

---

*These ordinary 4-D vectors aren't the invariant "four-vectors" of standard relativity theory.

In terms of our 3-D reference space, **A**'s components are

$$A_x = v_x \tag{3-19}$$

$$A_y = v_y \tag{3-20}$$

$$A_z = v_z \tag{3-21}$$

$$A_\phi = \sqrt{c^2 - (v_x^2 + v_y^2 + v_z^2)} \tag{3-22}$$

where **v** refers to the 3-D velocity of the incident pro-particle.

Similarly, **C**'s components are

$$C_x = v_{xr} \tag{3-23}$$

$$C_y = v_{yr} \tag{3-24}$$

$$C_z = v_{zr} \tag{3-25}$$

$$C_\phi = \sqrt{c^2 - (v_{xr}^2 + v_{yr}^2 + v_{zr}^2)} \tag{3-26}$$

where the r's in subscripts indicate the "reflected" pro-particle. (Later we'll switch to the usual optical notation, where reflected rays are indicated with primes. However, in the relativistic equations we'll be deriving here, primes signify moving reference frames.)

While 4-D vectors will be used to guide our analysis, what we're after are equations describing pro-particle reflection in 3-D terms. Thus we'll be going back and forth between the above 4 and 3-D velocity representations.

To find vector **B**, we need an equation for the mirror/wavefront contact hyperplane. This can be found by solving simultaneous equations for the mirror and incident wavefront. We'll start by writing the wavefront equation. In our 4-D space, the normal form for the equation of a fixed hyperplane is

$$x \cos\alpha + y \cos\beta + z \cos\gamma + \phi \cos\delta = p \tag{3-27}$$

where p is the perpendicular distance from the origin, O, to the hyperplane at point P, and $\alpha$, $\beta$, $\gamma$, $\delta$ are the angles between the perpendicular, OP, and the positive x, y, z, $\phi$ axes respectively.

If we take our wavefront to cross the origin at time t = 0, the perpendicular distance, p, from the origin to the wavefront, is given by ct. Since vector **A** is normal to the incident wavefront, the direction cosines of a perpendicular are given by **A**'s components divided by its magnitude, c. Putting all these into Eq. (3-27), we have an equation for the wavefront as a function of time:

$$x\left(\frac{A_x}{c}\right) + y\left(\frac{A_y}{c}\right) + z\left(\frac{A_z}{c}\right) + \phi\left(\frac{A_\phi}{c}\right) = ct \qquad (3\text{-}28)$$

In terms of 3-D velocities, this is:

$$x\left(\frac{v_x}{c}\right) + y\left(\frac{v_y}{c}\right) + z\left(\frac{v_z}{c}\right) + \phi\left(\frac{\sqrt{c^2 - (v_x^2 + v_y^2 + v_z^2)}}{c}\right) = ct \qquad (3\text{-}29)$$

Again, the mirror is perpendicular to x and moves along that axis with velocity w. If it crosses the origin at time t = 0, the mirror equation is simply

$$x = wt \qquad (3\text{-}30)$$

or

$$t = \frac{x}{w} \qquad (3\text{-}31)$$

Putting this expression for t into Eq. (3-29) gives a time-independent equation for the mirror/wavefront contact hyperplane:

$$x\left(\frac{v_x}{c}\right) + y\left(\frac{v_y}{c}\right) + z\left(\frac{v_z}{c}\right) + \phi\left(\frac{\sqrt{c^2 - (v_x^2 + v_y^2 + v_z^2)}}{c}\right) = \frac{cx}{w} \qquad (3\text{-}32)$$

Transposing the term on the right and multiplying by a constant, k, we can write

$$x\left(k\frac{v_x}{c} - k\frac{c}{w}\right) + y\left(k\frac{v_y}{c}\right) + z\left(k\frac{v_z}{c}\right) + \phi\left(k\frac{\sqrt{c^2 - (v_x^2 + v_y^2 + v_z^2)}}{c}\right) = 0 \qquad (3\text{-}33)$$

Next, we'll put this hyperplane equation into normal form. In so doing, the coordinate coefficients will give us the direction cosines for vector **B**. In the normal form, the sum of the squared coordinate coefficients equals one. We can satisfy this condition with an appropriate value for k. Setting the squared coefficients equal to one, we have

$$k^2 \left[ \left( \frac{v_x}{c} - \frac{c}{w} \right)^2 + \left( \frac{v_y}{c} \right)^2 + \left( \frac{v_z}{c} \right)^2 + \left( \frac{\sqrt{c^2 - (v_x^2 + v_y^2 + v_z^2)}}{c} \right)^2 \right] = 1 \qquad (3\text{-}34)$$

Expanding and collecting terms gives

$$k^2 \left( 1 - \frac{2v_x}{w} + \frac{c^2}{w^2} \right) = 1 \qquad (3\text{-}35)$$

The normalizing values for $k^2$ and k then are

$$k^2 = \frac{1}{1 - \frac{2v_x}{w} + \frac{c^2}{w^2}} \qquad (3\text{-}36)$$

and

$$k = \frac{1}{\pm\sqrt{1 - \frac{2v_x}{w} + \frac{c^2}{w^2}}} \qquad (3\text{-}37)$$

(In the end, we'll be using $k^2$ instead of k, so the sign of k doesn't matter.) With this k, the coordinate coefficients in Eq. (3-33) are the direction cosines for vector **B**. Multiplied by **B**'s magnitude, c, they give its components:

$$B_x = k \left( v_x - \frac{c^2}{w^2} \right) \qquad (3\text{-}38)$$

$$B_y = k v_y \qquad (3\text{-}39)$$

$$B_z = k v_z \qquad (3\text{-}40)$$

$$B_\phi = k \sqrt{c^2 - (v_x^2 + v_y^2 + v_z^2)} \qquad (3\text{-}41)$$

Now that we have vector **B**, we'll find the cosine of the angle, θ, between -**A** and **B**. Since both these vectors have magnitude c, cosθ is given by

$$\cos\theta = \frac{-\mathbf{A} \cdot \mathbf{B}}{c^2} \qquad (3\text{-}42)$$

From our equations for the components of **A** and **B** we have

$$-\mathbf{A}\cdot\mathbf{B} = -A_x B_x - A_y B_y - A_z B_z - A_\phi B_\phi$$

$$= k\left(-v_x^2 + \frac{c^2 v_x}{w}\right) - kv_y^2 - kv_z^2 - k[c^2 - (v_x^2 + v_y^2 + v_z^2)] \qquad (3\text{-}43)$$

$$= k\left(\frac{c^2 v_x}{w} - c^2\right)$$

From the prior equation, dividing this by $c^2$ gives the cosine of $\theta$:

$$\cos\theta = k\left(\frac{v_x}{w} - 1\right) \qquad (3\text{-}44)$$

Now we can write equations for the components of vector C, using Eq. (3-17), our equations for the components of A and B, and that for the cosine of $\theta$. For the x component, we have

$$C_x = A_x + 2B_x \cos\theta$$

$$= v_x + 2k\left(v_x - \frac{c^2}{w}\right)k\left(\frac{v_x}{w} - 1\right) \qquad (3\text{-}45)$$

$$= v_x + 2k^2\left(\frac{v_x^2}{w} - v_x - \frac{c^2 v_x}{w^2} + \frac{c^2}{w}\right)$$

To simplify this, we write an expression equal to $v_x$, having several terms. Starting with $v_x$, we divide and multiplying by $k^2$, then substitute the expression for $k^2$ from Eq. (3-36) into the denominator:

$$v_x = k^2 v_x / k^2$$

$$= k^2 v_x\left(1 - \frac{2v_x}{w} + \frac{c^2}{w^2}\right) \qquad (3\text{-}46)$$

$$= k^2\left(-\frac{2v_x^2}{w} + v_x + \frac{c^2 v_x}{w^2}\right)$$

Substituting this expression for $v_x$ for the first term in Eq. (3-45) gives

$$C_x = k^2\left(-v_x - \frac{c^2 v_x}{w^2} + \frac{2c^2}{w}\right) \qquad (3\text{-}47)$$

From the same group of equations, we can write one for the y component of vector C. We have:

$$C_y = A_y + 2B_y \cos \theta$$

$$= v_y + 2k(v_y)k\left(\frac{v_x}{w} - 1\right) \quad (3\text{-}48)$$

$$= v_y + 2k^2\left(\frac{v_x v_y}{w} - v_y\right)$$

This can be simplified by the previous method. Multiplying and dividing $v_y$ by $k^2$, then substituting for $k^2$ gives

$$v_y = k^2 v_y / k^2$$

$$= k^2 v_y \left(1 - \frac{2v_x}{w} + \frac{c^2}{w^2}\right) \quad (3\text{-}49)$$

$$= k^2 \left(\frac{c^2 v_y}{w^2} - \frac{2v_x v_y}{w} + v_y\right)$$

Substituting this expression for $v_y$ for the first term in Eq. (3-48) yields

$$C_y = k^2 \left(\frac{c^2 v_y}{w^2} - v_y\right) \quad (3\text{-}50)$$

Similarly, for $C_z$ we get

$$C_z = k^2 \left(\frac{c^2 v_z}{w^2} - v_z\right) \quad (3\text{-}51)$$

While it won't be used here, we'll also give the expression for $C_\phi$:

$$C_\phi = A_\phi + 2B_\phi \cos \theta$$

$$= \sqrt{c^2 - (v_x^2 + v_y^2 + v_z^2)} + 2k\sqrt{c^2 - (v_x^2 + v_y^2 + v_z^2)}\, k\left(\frac{v_x}{w} - 1\right) \quad (3\text{-}52)$$

$$= \left(1 + \frac{2k^2 v_x}{w} - 2k^2\right)\sqrt{c^2 - (v_x^2 + v_y^2 + v_z^2)}$$

Now we'll rewrite Eqs. (3-47), (3-50), and (3-51), putting them entirely in terms of 3-D pro-particle velocities and the mirror velocity, w. For $k^2$, we substitute the expression from Eq. (3-36). Rewriting Eq. (3-47), we get

$$v_{xr} = \frac{-v_x - c^2 v_x/w^2 + 2c^2/w}{1 - 2v_x/w + c^2/w^2} \tag{3-53}$$

The simplest form of these equations would be obtained by multiplying top and bottom by $w^2$. However, for deriving standard relativistic equations, a more useful form is obtained by multiplying top and bottom by $w^2/c^2$. Doing so gives*

$$v_{xr} = \frac{-v_x - v_x w^2/c^2 + 2w}{1 - 2v_x w/c^2 + w^2/c^2} \tag{3-54}$$

Following the same procedure with Eqs. (3-50) and (3-51) gives

$$v_{yr} = \frac{v_y - v_y w^2/c^2}{1 - 2v_x w/c^2 + w^2/c^2} \tag{3-55}$$

and

$$v_{zr} = \frac{v_z - v_z w^2/c^2}{1 - 2v_x w/c^2 + w^2/c^2} \tag{3-56}$$

These last three equations, describing pro-particle reflection in 3-D terms, are the set we've been chasing.

---

*If both sides of this equation are divided by c, it also describes the relativistic reflection of light from a mirror moving in x. When this is done, $v_x/c$ and $v_{xr}/c$ are the direction cosines for incident and reflected light rays in, for example, the x-y plane. This equation is given in Einstein's 1905 "On the Electrodynamics of Moving Bodies". However, unlike Einstein, who derived it from a "principle" of relativity, here we've done so from a physical system, without assuming this principle.

# Relativistic Velocity Transformation

From our pro-particle reflection equations, now we can derive those for the relativistic transformation of velocity. These are:

$$u_x = \frac{u_x' + v}{1 + u_x' v / c^2} \tag{3-57}$$

$$u_y = \frac{u_y' \sqrt{1 - v^2/c^2}}{1 + u_x' v / c^2} \tag{3-58}$$

$$u_z = \frac{u_z' \sqrt{1 - v^2/c^2}}{1 + u_x' v / c^2} \tag{3-59}$$

where u is a velocity observed in some chosen "rest" frame, and u' is the corresponding velocity measured within a "moving" reference frame, with a relative velocity, v, in the x dimension. (While the previously derived Lorentz equations transform coordinates from a "rest" frame to a frame in relative motion, these equations go the opposite way. This form was chosen here because Eq. (3-57) is the familiar "addition of velocities" equation.)

Our approach will be to compare the final velocities of the same incident pro-particle after reflection by a fixed mirror and by a moving one. We'll see that the effect of mirror motion is to transform the reflected velocity according to Eqs. (3-57)-(3-59) above. The (invariant) velocity, u', will correspond to a fixed mirror, while the transformed velocity, u, will result from a moving one. Also, we'll see that the reference frame velocity, v, is a function of the mirror velocity, w.

To describe the reflection of a pro-particle by a fixed mirror, we can put a zero value for the mirror velocity, w, in Eqs. (3-54)-(3-56). In this case, of course the incident and reflected velocities are equal, except that the x component is reversed. So, calling this reflected velocity u', we can write

$$-u_x' = v_x \tag{3-60}$$

$$u_y' = v_y \tag{3-61}$$

$$u_z' = v_z \tag{3-62}$$

To describe the same incident pro-particle, reflected by a moving mirror, we can substitute these expressions into Eqs. (3-54)-(3-56). Calling the reflected velocity in this case u, we have

$$u_x = \frac{u'_x + u'_x w^2/c^2 + 2w}{1 + 2u'_x w/c^2 + w^2/c^2} \quad (3\text{-}63)$$

$$u_y = \frac{u'_y - u'_y w^2/c^2}{1 + 2u'_x w/c^2 + w^2/c^2} \quad (3\text{-}64)$$

$$u_z = \frac{u'_z - u'_z w^2/c^2}{1 + 2u'_x w/c^2 + w^2/c^2} \quad (3\text{-}65)$$

where u' is the (invariant) velocity of a pro-particle reflected by a fixed mirror, and u is that of the same incident pro-particle reflected by a mirror with velocity, w.

If we put a zero value for u' in Eqs. (3-63)-(3-65), the resulting velocity, u, has no y or z components and is given entirely by Eq. (3-63). This reduces to

$$u_x = \frac{2w}{1 + w^2/c^2} \quad (3\text{-}66)$$

For a pro-particle which is stationary after "reflection" from a fixed mirror, this gives the velocity of the same incident pro-particle after reflection from a moving one. If the effect of mirror motion is a relativistic transformation of velocity, then this pro-particle should be stationary once again in some new relativistic frame. In other words, its velocity would correspond to that of this new frame. This gives

$$v = \frac{2w}{1 + w^2/c^2} \quad (3\text{-}67)$$

where v is the reference frame velocity corresponding to a given mirror velocity, w. (Notice the similarity to Eq. (3-57).)

In the case of a pro-particle with an arbitrary value of u', the transformed velocity, u, should have a proper relativistic value in the reference frame specified by the last equation. If we can demonstrate this, it would establish that our moving mirror transform is relativistic for reflected pro-particles in general.

Rearranging Eq. (3-63), then dividing numerator and denominator by $(1 + w^2/c^2)$, we can write

$$u_x = \frac{u_x'(1+w^2/c^2) + 2w}{(1+w^2/c^2) + 2u_x'w/c^2}$$

$$= \frac{u_x' + \left(\dfrac{2w}{1+w^2/c^2}\right)}{1 + u_x'\left(\dfrac{2w}{1+w^2/c^2}\right)/c^2} \quad (3\text{-}68)$$

From Eq. (3-67), we then replace the terms in parentheses by v, to obtain

$$u_x = \frac{u_x' + v}{1 + u_x'v/c^2}$$

which is the first relativistic velocity transform equation, Eq. (3-57).

Now we'll derive the second velocity transform equation. Rearranging Eq. (3-64), then dividing top and bottom by $(1 + w^2/c^2)$, we have

$$u_y = \frac{u_y'(1-w^2/c^2)}{(1+w^2/c^2) + 2u_x'w/c^2}$$

$$= \frac{u_y'\left(\dfrac{1-w^2/c^2}{1+w^2/c^2}\right)}{1 + u_x'\left(\dfrac{2w}{1+w^2/c^2}\right)/c^2} \quad (3\text{-}69)$$

Here, we put the latter coefficient of $u_y'$ into a different form:

$$\frac{1-w^2/c^2}{1+w^2/c^2} = \sqrt{\frac{(1-w^2/c^2)^2}{(1+w^2/c^2)^2}}$$

$$= \sqrt{\frac{(1+w^2/c^2)^2 - 4w^2/c^2}{(1+w^2/c^2)^2}} \quad (3\text{-}70)$$

$$= \sqrt{1 - \left(\frac{2w}{1+w^2/c^2}\right)^2/c^2}$$

Substituting this expression for the coefficient of $u_y'$ in the second part of Eq. (3-69) gives

$$u_y = \frac{u_y' \sqrt{1 - \left(\frac{2w}{1+w^2/c^2}\right)^2 / c^2}}{1 + u_x' \left(\frac{2w}{1+w^2/c^2}\right)/c^2} \tag{3-71}$$

From Eq. (3-67), we can then substitute v for the expressions in parentheses to obtain

$$u_y = \frac{u_y' \sqrt{1 - v^2/c^2}}{1 + u_x' v / c^2}$$

which is Eq. (3-58). An identical procedure also gives Eq. (3-59), for $u_z$.

Here we've seen that, using Eq. (3-67) as the velocity of our moving reference frame, all of the relativistic velocity transform equations are obtained. Again, this shows that the 3-D velocity of *any* reflected pro-particle (not just an initially motionless one) is transformed to its proper value for this moving frame.

As noted previously, reflection from a moving hyperplane mirror tilts wavefronts without changing the 4-D wavefront length between pro-particles. Earlier, we found that, given this condition, we could derive the first three Lorentz equations for the transformation of space. From those equations, plus these just obtained for velocity, the last Lorentz transform equation, for time, can be derived. (This is just a reversal of the usual derivation of the velocity transform from that for space and time.) So, this "moving mirror" transform is a complete one.

Also, in our initial derivation of the Lorentz transform, we only considered pairs of pro-particles, with identical velocities, associated with the same hyperplane waves. As derived here, the transform applies to any ensemble of reflected pro-particles, having any combination of velocities. Thus, if everything were somehow made of such pro-particles, we'd have a relativistic system. Later, when we arrive at more realistic particles, we'll find that, since they exist within the same wave system, they inherit the basic relativistic behavior of these reflected pro-particles.

# 4. ϕ Flows and the Scalar Potential

So far, we've been treating the waves in this system as though they might be linear ones. However, as mentioned earlier, this theory is based on flow waves, and these are inherently nonlinear. As a first step toward exploring their nonlinear interactions, here we'll look at interactions of steady flows and waves. In particular, we'll be examining flows directed along the ϕ dimension, and their effects on the motion of pro-particles. We'll see that these effects are identical to those of stationary scalar electric potentials on charged particles.*

Just as sound is reflected and refracted by a stream of moving air, so too are the waves in this system, when they encounter flows. In analyzing these effects, we'll use the approach taken for reflection by a moving mirror: First we'll use Huygens' principle to derive a law -- of refraction here -- then we'll apply it to the behavior of pro-particles. We'll be working in just two dimensions again, for simplicity, instead of the full 4-dimensional space. We'll also change to optical notation, using primes now to indicate quantities "after refraction."

**Fig. 4-1.** Wave refraction by a flowing medium. The medium on the left is fixed, while in the shaded area it moves downward with velocity U'. As shown in (a), the effect here is to decrease the angle, θ', between the wavefront and the flow interface. Part (b) shows the corresponding refraction of a ray.

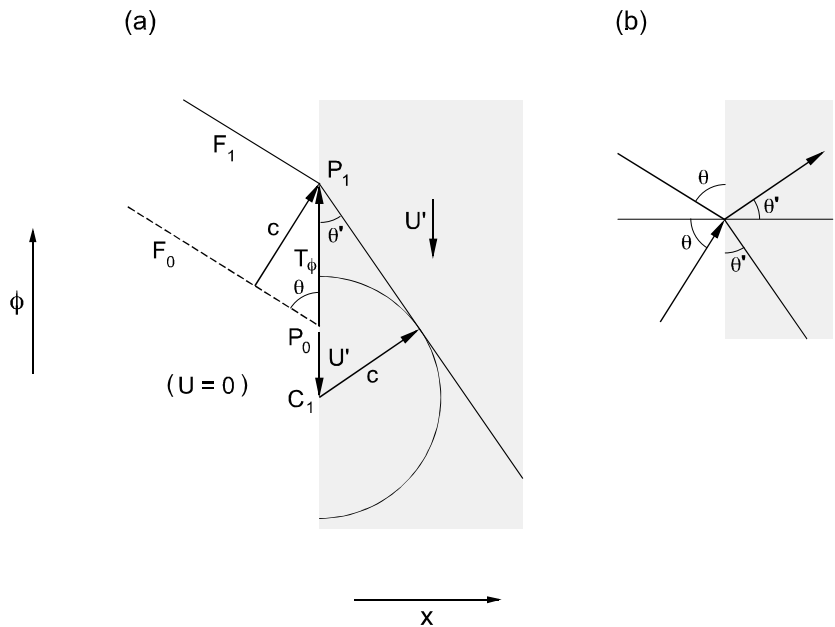

---

*A very insightful introduction to the scalar and vector potentials is Feynman, Leighton, and Sands[17].

Fig. 4-1(a) depicts a wavefront entering a flow region.  While the wave medium (stuff) has no overall flow on the left, on the right it flows directly downward with velocity U' in the φ dimension.  (As in ordinary fluids, there is always a continuous gradient between different flows.  However, we'll assume that the gradient width here is arbitrarily small, so we have a sharp interface.)  The broken line, $F_0$, depicts just the incident portion of the wavefront, which is moving up and to the right, at time t = 0.  The bent, solid line, $F_1$, shows the whole wavefront, including the refracted part, at a later time, t = 1.

As usual, vectors with magnitude c are rays, normal to the wavefront.  Points $P_0$ and $P_1$ are the successive intersection points between the wavefront and interface.  The semicircle to the right of $P_0$ represents a Huygens wavelet which originated at this point at t = 0.  As depicted by the vector U', the wavelet is displaced by the flow, so at t = 1 its center is located at $C_1$. (Like the circular ripples from a tossed stone, carried downstream by a river.)  By Huygens' principle, the refracted wavefront passes through point $P_1$ and is tangent to this wavelet.  The effect here is to decrease the angle θ' between the refracted wavefront and the interface.

To write a refraction equation, we can start by observing that, in terms of a fixed coordinate system, the overall phase velocity of the wavefront must be the same along both sides of the interface.  From this, we can write

$$V_\phi = T_\phi + U = T_\phi' + U' \tag{4-1}$$

where $V_\phi$ is the φ phase velocity with respect to the reference space, $T_\phi$ is the φ phase velocity with respect to the wave medium, U is the φ velocity of the medium, and primes indicate "after refraction."

From the left side of the Fig. 4-1(a), you can see that the cosecant of θ is given by

$$\csc\theta = \frac{T_\phi}{c} \tag{4-2}$$

where θ is the angle between the incident wavefront and interface.  Substituting for $T_\phi$ and $T_\phi'$ in the previous equation, we can write

$$c \cdot \csc\theta' + U' = c \cdot \csc\theta + U \tag{4-3}$$

For the refracted wave angle, this gives

$$\csc\theta' = \csc\theta + (U - U')/c \tag{4-4}$$

As depicted in Fig. 4-1(b), θ and θ' are also the angles of incident and refracted rays with respect to an interface normal.  So here we have the equivalent of Snell's (Descartes') law for ordinary refraction.

**Fig. 4-2.** Unlike Fig. 4-1, here the φ component of the wave velocity and the flow, U', are in the same direction. In this case θ' is increased.

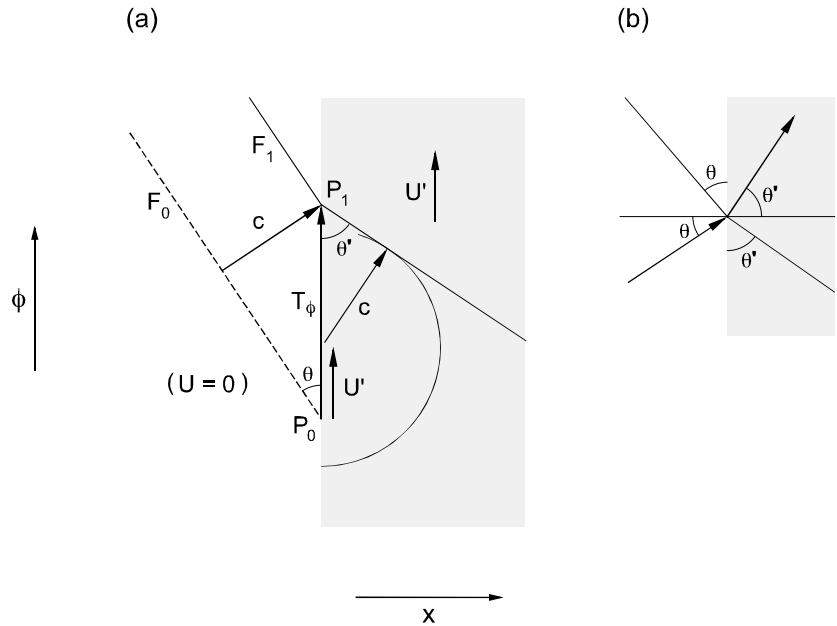

We saw in Fig. 4-1 that, when a refracting flow runs opposite to the φ component of incident rays, the rays are bent *toward* an interface normal. Fig. 4-2 shows the other case, where the flow is in the same φ direction as the incident rays. In this case, the refracted rays are deflected *away* from an interface normal. With increasing flow in this direction, a point is eventually reached where the refracted rays run parallel to the interface and no longer enter the refracting medium. As with ordinary refraction, this condition marks the onset of total reflection.

At the critical angle for total reflection, we have θ' = 90 degrees, or a cosecant of 1. Substituting this value for csc θ' in Eq. (4-3) and rearranging, we find that the limiting value for U' is given by

$$U' = U + c(\csc\theta - 1) \qquad (4\text{-}5)$$

For a positive θ, a larger U' results in total reflection. The angles of reflected waves are given by the usual Huygens' construction. So, with a *stationary* interface, there is the familiar law of reflection, with equal angles of incidence and reflection.

(The preceding equations have been known for some time. They can be found in Rayleigh's 1896 treatise, *The Theory of Sound*[17] [still in print], which has a section on the interaction of wind and sound. Rayleigh was particularly concerned with the observation that sound can carry downwind for surprising distances. He credits an 1857 paper by Stokes as first pointing out the role of wind refraction in this effect. Rayleigh also gives an interesting account of various experiments on the effects of wind on sound. Once recognized experimental physics could be done just by ringing a bell in an empty field, noting the wind, and listening for the sound at different locations!)

Although our refraction equations are accurate, I'd like to mention that the physical picture on which they are based is somewhat oversimplified. Ribner[18] has pointed out that a straight interface between two sharply divided flows is necessarily distorted by the pressures of impinging acoustic waves. Ribner treats the distorted interface as a "wavy wall." The ripples in the wall interact with the flows moving past it on either side and generate additional wave components. (In the context of our wave system, these ripples would have a ϕ period corresponding to the structure interval, s.) Of course the ϕ phase velocity of the ripples is the same as that of the incident waves which cause them. The result is that the additional components created have the same angles as the basic refracted and reflected waves described above.*

Although the wave angles are no different with this effect, the predicted amplitudes change significantly. What's involved is the transfer of energy between the flows and waves. Thus the waves are nonlinear. Also, in terms of the incident wave amplitude, the effect is first order, so it can't be neglected. (It would be interesting to know its role in downwind sound propagation.) Ribner gives corrected equations for the refracted and transmitted amplitudes. The condition for total reflection remains Eq. (4-5). However, at this point we won't be concerned with amplitudes.

**Fig. 4-3.** Positive and negative pro-particles, with the same velocity and associated wavelength in the reference space.

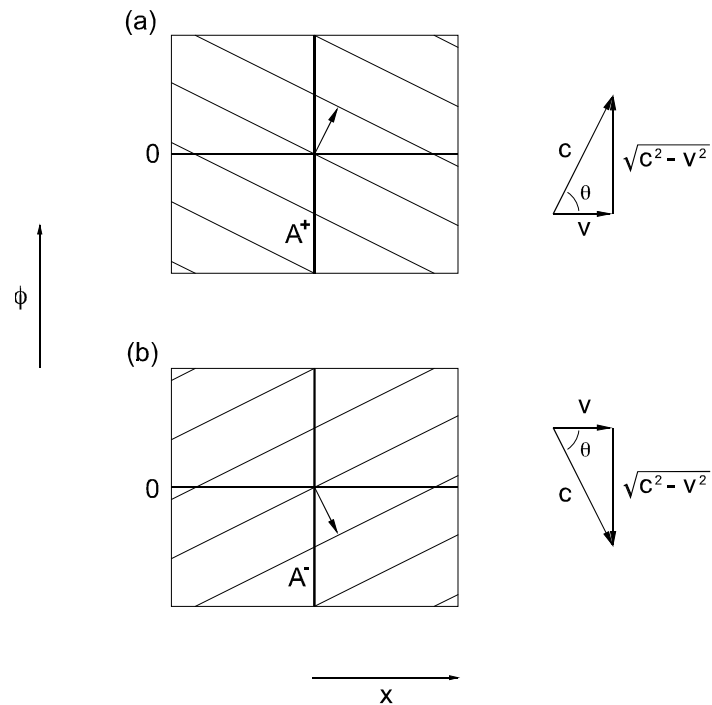

---

*This assumes that the flow isn't changed significantly by its interaction with the waves. To the degree that the flow velocity does change, of course there *would* be a change in the refracted wave angle.

Returning to our repetitive wave system, now we'll look at the refraction of pro-particles by ϕ flows. We'll start by observing that different pro-particles, with opposite ϕ velocity components, can have the same velocity in the reference space. This is illustrated in Fig. 4-3. The waves in part (b) are a mirror image of those in (a), having the same wavelength and phase velocity in the reference space. A⁺ and A⁻ represent similar pro-particles associated with these waves. Both are moving in the positive x direction with the same velocity, v. But in the ϕ dimension, A⁺ is moving positively, while A⁻ is going the opposite way. To differentiate these cases, we'll classify pro-particles by the sign of their ϕ velocity component.

For example, taking the ray in Fig. 4-1(b) to represent a pro-particle trajectory, this figure shows a positive pro-particle entering a negative ϕ flow. In this case, the x component of the ray, corresponding to the pro-particle's velocity in the reference space, increases. Fig. 4-2(b) depicts a positive pro-particle entering a negative flow. Here the x component decreases. It follows that, *when a pro-particle enters a ϕ flow of the same sign, its 3-D velocity decreases. Or, if it enters a flow of the opposite sign, the velocity increases*. This behavior is analogous to that of free charged particles entering different electric potentials. As we'll see, the analogy is an exact one.

Next we'll rewrite our refraction equations in terms of pro-particle velocities. In Eq. (4-3), the orientations of rays are given as the cosecant of the angle θ with an interface normal. In the 2-D case we're working with, this normal parallels the x axis. As illustrated by the vector triangle in Fig. 4-3(a), the cosecant of θ for a positive pro-particle is given by

$$\csc \theta = \frac{c}{\sqrt{c^2 - v^2}} = \frac{1}{\sqrt{1 - v^2/c^2}} \qquad (4\text{-}6)$$

For the general case, we can write

$$\csc \theta = \frac{1}{Q\sqrt{1 - v^2/c^2}} \qquad (4\text{-}7)$$

where we define Q as +1 for a positive pro-particle, and -1 for a negative type.

(The vector triangles of Fig. 4-3 give the pro-particle velocity components with respect to the wave medium. When the medium is flowing in the ϕ dimension, of course the ϕ velocity component has a different value with respect to fixed coordinates. Nevertheless, v, the x velocity component, is the same in either coordinate system.)

Putting this expression for the cosecant of θ into both sides of Eq. (4-3), then multiplying by Q gives

$$\frac{c}{\sqrt{1 - v'^2/c^2}} + QU' = \frac{c}{\sqrt{1 - v^2/c^2}} + QU \qquad (4\text{-}8)$$

If we set U equal to zero (the case shown in Figs. 4-1 and 4-2), and solve for the refracted velocity, v', we get

$$v' = \pm c \sqrt{1 - \left(\frac{1}{\frac{1}{\sqrt{1-v^2/c^2}} - \frac{QU'}{c}}\right)^2} \qquad (4\text{-}9)$$

where the sign of the right side corresponds to that of v.

Now we'll compare this to the velocity of a charged particle entering a uniform scalar electric potential. Since the above pro-particle moves only in the x reference space dimension, we'll look at an analogous charged particle moving in x only, across a sharp potential boundary. From the relativistic conservation of energy for a free particle, we can write

$$E = \frac{m_0 c^2}{\sqrt{1-v^2/c^2}} + q\Phi = \frac{m_0 c^2}{\sqrt{1-v'^2/c^2}} + q\Phi' \qquad (4\text{-}10)$$

where E is the energy, $m_0$ is the particle's rest mass, q its charge, $\Phi$ the scalar electric potential, and primes denote "after the boundary." Dividing by $m_0 c$, we can write another conservation equation resembling Eq. (4-8):

$$\frac{c}{\sqrt{1-v'^2/c^2}} + \frac{q\Phi'}{m_0 c} = \frac{c}{\sqrt{1-v^2/c^2}} + \frac{q\Phi}{m_0 c} \qquad (4\text{-}11)$$

In Eq. (4-8), if we scale the flow, U, such that

$$QU = \frac{q\Phi}{m_0 c} \qquad (4\text{-}12)$$

it becomes equivalent to Eq. (4-11). Also, if we put this expression for QU into Eq. (4-9), it gives

$$v' = \pm c \sqrt{1 - \left(\frac{1}{\frac{1}{\sqrt{1-v^2/c^2}} - \frac{q\Phi'}{m_0 c^2}}\right)^2} \qquad (4\text{-}13)$$

which is the solution of Eq. (4-11) for $\Phi = 0$. So, given the condition of Eq. (4-12), our pro-particle's velocities behave *identically* to those of real particles with charge q and mass $m_0$, with $\phi$ flows playing the role of scalar potentials.

We can pursue this analogy a little further. We started with the conservation of two different quantities: pro-particle ϕ phase velocity, and the energy of a free charged particle. Note that energy of a particle and its de Broglie wave frequency, ν, are related by

$$E = h\nu \qquad (4\text{-}14)$$

where h is Planck's constant. So there is also a conserved frequency. From Eq. (4-10), this can be written as

$$\nu = \frac{1}{h}\left(\frac{m_0 c^2}{\sqrt{1-v^2/c^2}} + q\Phi\right) \qquad (4\text{-}15)$$

In our repetitive wave system, the conservation of ϕ phase velocity means there is also a conserved frequency for pro-particles. Dividing the ϕ phase velocity represented by Eq. (4-8) by the ϕ structure interval, s, we can write

$$\nu = \frac{1}{s}\left(\frac{c}{\sqrt{1-v^2/c^2}} + QU\right) \qquad (4\text{-}16)$$

calling this frequency ν also. In Section 2, we found that, with s equal to $h/m_0 c$, the wavelengths associated with pro-particles match those of real particles with mass $m_0$. Taking this value for s again, the last equation becomes

$$\nu = \frac{m_0 c}{h}\left(\frac{c}{\sqrt{1-v^2/c^2}} + QU\right) \qquad (4\text{-}17)$$

If we also substitute for QU from Eq. (4-12), this gives Eq. (4-15), the wave frequency of a real particle in a scalar potential. So here we find another fundamental similarity between the waves of this new system and those of quantum mechanics.

We've seen now that pro-particles can precisely model various features of particles having a single rest mass, $m_0$. Of course, in nature, not all particles have the same mass. Another apparent shortcoming of this particle model is that, as ray analogues, pro-particles can be partially refracted and reflected. Of course, when real particles, such as electrons, meet sharp changes in the scalar potential, they are either "refracted" or reflected; the whole particle goes one way or the other.

Pro-particles are just a construct (wave markers) that we've used to describe some fundamental characteristics of the waves in this system. In a subsequent section, more complex wave-based entities will be introduced, which resemble real particles much more closely. However, because they exist within the same relativistic wave system, they may inherit much of the behavior of our provisional particles -- including their velocity behavior in ϕ flows. The realistic particle model assumes that ϕ flows are, in fact, the physical basis of the scalar electric potential.

# 5. Flow Waves and the Vector Potential

In the last section, we found a correspondence between the behavior of pro-particles in stationary ϕ flows, and that of real charged particles in scalar potentials. Taking a similar approach, here we'll describe pro-particles entering *moving* ϕ flows. This will be compared to real particles entering moving scalar/vector potentials. In so doing, we'll introduce the vector potential, while further defining the nature of the scalar potential in this theory.

A basis for moving ϕ flows is provided by another feature of our wave system: its flows are not steady ones, but are due to acoustic flow waves. (So, while flows influence the propagation of waves, these also consist of waves themselves.) We'll see that the lateral movement of ϕ flows is made possible by this underlying wave nature. To introduce acoustic flow waves, we'll begin by looking at the familiar example of sound waves in air.

A common device for describing sound wave generation is an air-filled cylinder with a piston. Suppose we have a very long cylinder, aligned with the x axis, and a piston which makes waves that propagate in the +x direction. If the piston oscillates around a fixed point (like a stereo speaker), there is no net movement of the air in the cylinder and we have what are regarded as ordinary sound waves.

What happens if the piston does *not* oscillate around a fixed point? For example, suppose it starts abruptly, moves some distance in the positive x direction, and stops abruptly. Here, a rectangular flow wave pulse is created, with a density greater than the surrounding air. As it moves down the cylinder, the effect of this pulse is to displace individual air particles, on average, by the same distance traveled by the piston. (Otherwise, near the piston, we would see a lasting, local increase in the air density and pressure.) This pulse represents *a moving region of flow*, in which the average velocity of the air particles matches that of the moving piston.

By repeatedly moving the piston, only in the positive x direction, and stopping it, we can also generate a series of such pulses. Like the water solitons in Fig. 1-5, the waves then would carry an intermittent, but continuing, flow.* Likewise, by repeatedly moving and stopping the piston in the negative x direction, we can generate a train of pulses (moving positively in x) with negative relative density. Note that, in this last case, the net flow is *opposite* the wave movement.

"Ordinary" acoustic waves can be viewed as a special case of flow waves. These involve flow wave components of both positive and negative relative density, where these components are exactly equal in magnitude and have identical velocity vectors.

---

*This progressive driving (as opposed to oscillating) mode of wave generation is common in nature. For example, this is essentially how sound is produced when you speak or play a wind instrument.

As with steady flows, other waves are refracted and reflected by flow waves. A simplified example is illustrated in Fig. 5-1. Parallel lines in the region on the right represent flow waves, moving vertically in the ϕ dimension. An average downward flow, represented by the vector U' is associated with these. Here we'll assume that these waves are such that the fluid density inside the flow wave region is the same as that outside it. (We can meet this condition as described below.) Incident on the flow wave region is a wavefront associated with longer-wavelength, small-amplitude waves. As indicated by the vector c, this is moving up and to the right.

**Fig. 5-1.** Refraction and reflection by relatively short-wavelength flow waves. Solid lines represent positive density waves, moving in the -ϕ direction, while broken lines mark negative density waves, going the opposite way. Vector U' indicates the average net flow.

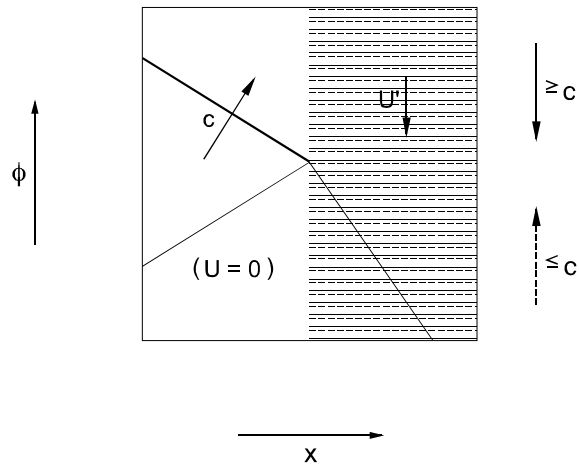

We know that, despite the chaotic motion of air particles at the molecular level, at larger scales, sound wave behavior is given by the *average* motions of air particles. From this, we can infer that, if the scale of our flow waves is relatively minuscule, the refraction and reflection of larger-scale waves should be determined by only the average flow, U'. (Also, with very short-wavelength flow waves, we can neglect diffraction at the boundary.) So the figure just depicts the usual refracted and reflected of wavefronts resulting from such a flow. *This illustrates a nonlinear wave effect which is independent of the wave medium. Instead, it depends entirely on a property of the waves themselves: their associated flow.*

Notice that the downward flow, U', in Fig. 5-1 could be realized by either positive density waves (solid lines) traveling downward or waves of negative relative density (broken lines) traveling upward. To meet our assumption that the fluid medium has the same average density inside the flow wave region as outside, these must be a balanced combination of both types. (For example, if we had positive waves only, the net density in the flow wave region would be greater.) As with "ordinary" acoustic waves, here we have equal positive and negative density components. But in this case, they are moving oppositely.

Referring again to a very long, air-filled cylinder, we can visualize the generation of such complementary flow waves. In this example, instead of having a piston, the ends of the cylinder are plugged. Here, waves are created inside the cylinder by moving the whole unit along the x axis. Suppose this cylinder is moved repeatedly, in the +x direction only, and stopped. At one end plug, waves of positive density are produced, traveling in the +x direction. Also, from the opposite end, there are corresponding pulses of negative density, going in the -x direction. With or without the waves, of course the net air density inside the closed cylinder remains the same.*

Fig. 5-2. Moving flow wave regions. The 3-D velocity of their left edges is v, that of associated pro-particles. Positive and negative density flow waves are depicted in (a) and (b) respectively, while (c) represents a balanced combination. In (c), the net average flow, $U_3'$, parallels the $\phi$ axis.

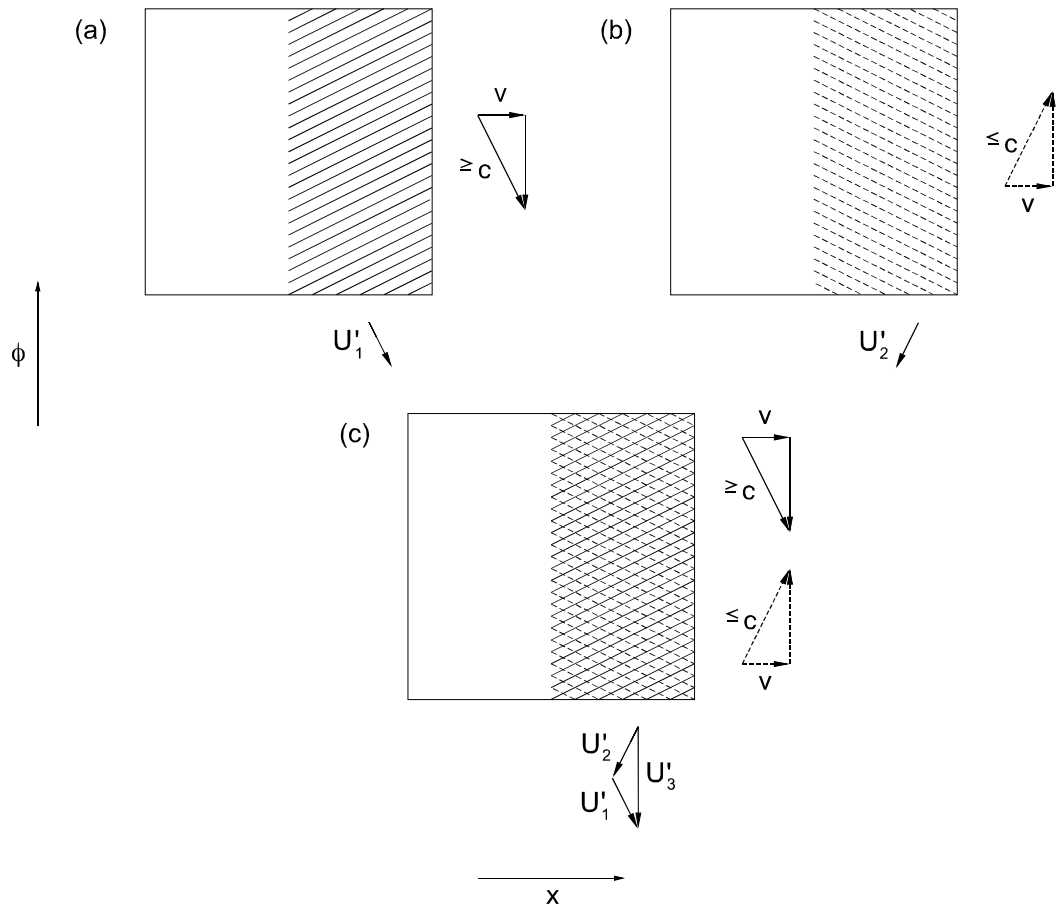

―――――――――――――

*Ship waves are another example of balanced flow waves. At the bow, where the hull displaces water, waves of positive net amplitude are produced. Although less noticeable, at the stern, waves of negative amplitude are also created. These bow and stern waves can move independently, with different velocities and trajectories. Since the overall ocean level doesn't change, of course the net amplitudes must be exactly opposite.

Here we return to the present theory, and the mechanism for the movement of localized $\phi$ flows. These move laterally in the x, y and z dimensions, without a corresponding migration of the surrounding fluid, stuff. For steady flows, this would seem impossible. However, as suggested in Fig. 5-2, such traveling flows can be realized on the basis of flow waves.

Fig. 5-2(a) shows a localized set of positive density flow waves, whose left edge is moving in the positive x direction. Again, in this simplified example, we'll assume their wavelength is arbitrarily short, so diffraction can be neglected. The edge's motion in the reference space corresponds to that of a pro-particle associated with the waves, represented by vector v in the vector triangle at the right. The vector $U_1'$, shown underneath, represents the average flow associated with the waves. Here this has the same direction as the wave motion, with components in the negative $\phi$ and positive x directions.

Another set of flow waves, with the same x velocity, v, appears in Fig. 5-2(b). These are moving to the right, like the last set, but also upward, in the opposite $\phi$ direction. As indicated by broken lines, these are waves of negative relative density. Here the associated flow, $U_2'$, is opposite the wave motion. The $\phi$ component of this flow is negative, as in (a), while the x flow component is opposite.

Fig. 5-2(c) shows a combination of two such wave sets. This represents balanced waves of positive and negative density, where again the average fluid density inside the wave region equals that outside it. Because they have the same velocity, v, in the reference space, the two wave sets travel together. The average net flow, $U_3$, is given by the vector sum of $U_1'$ and $U_2'$. The x components of the latter are exactly opposite, since these are the products of v and the average wave amplitudes. So, these cancel and $U_3'$ parallels the $\phi$ axis.

(Note that, while the wavefronts in Fig. 5-2(c) are depicted as straight, this may only be correct for small amplitudes. At larger amplitudes, these waves may distort one another, giving "wavy" fronts. Also, at large amplitudes, the 4-D component wave velocities may differ from c. Nevertheless, with appropriate tilts of the component waves, these should still move together in the reference space.)

This introduces another general quality of the wave system in this theory: *its flow waves are balanced, such that net flows only occur in the $\phi$ dimension.* (If not, it would follow that light or actual particles could be observed moving faster than c.) Below, we'll see that scalar/vector potentials correspond exactly to localized $\phi$ flows, moving in x, y and z, without net movements of the fluid itself in these dimensions.

While we've been primarily concerned with the refractive effects of $\phi$ flows, they also reflect waves, as illustrated in Fig. 5-1 for a stationary flow wave region. Earlier, in deriving the Lorentz transform, we assumed a reflection mechanism aligned with the $\phi$ axis. This was assumed to move through the wave medium without inducing a flow around the mirror. Notice that, *in these moving, wave-based $\phi$ flows, we have such a mechanism*.

How would the balanced waves of moving ϕ flow regions transform if they *themselves* were reflected by moving mirrors? For small amplitude waves, our previous "moving mirror" transform applies directly. Transforming a stationary flow wave region, such as that depicted in Fig. 5-1, gives a moving region like that shown in Fig. 5-2(c). As before, the 4-D length of hyperplane wavefront sections is unchanged, while the width of a moving wave region is contracted relativistically by the usual factor. Since they have opposite ϕ velocities, the positive and negative density component waves tilt oppositely with respect to the reference space.

It's assumed that, *in combination*, the flow waves of this system transform relativistically at larger amplitudes also. (Because they are incomplete, the simplified waves of Figs. 5-1 and 5-2 probably would not behave properly at large amplitudes.) Another important assumption is that each complementary pair of wavefronts crossing the reference space carries an invariant ϕ displacement of the wave medium. (As described later, it should be possible to check these assumptions by computer simulation.) *The average ϕ flow is then proportional to the flow wave frequency seen in the reference space*.

Given this, what is the transformed ϕ flow for a moving wave region? Calling the region velocity w now, at a fixed point in the reference space the flow wave frequency differs by the factor $1/\sqrt{1-w^2/c^2}$. Consequently, we can write the average ϕ flow seen at a fixed point as

$$U = U_0/\sqrt{1-w^2/c^2} \qquad (5\text{-}1)$$

where $U_0$ is the flow of a corresponding stationary wave region. So, while the region width contracts by the factor $\sqrt{1-w^2/c^2}$, the effective flow per unit width increases by the reciprocal, and total flow is conserved. In this respect, our wave-based flows behave the same as ordinary ones, and our assumption that the associated ϕ flow is proportional to the flow wave frequency seems justified.

Note that the wave frequency at a moving point in the reference space is dependant on its velocity. Consequently, the effective ϕ flow at such a point depends on its velocity also. (I.e., if more flow waves are encountered, the effective flow increases, and vice versa.) For example, at a point moving together with a flow wave region, the wave frequency is decreased by $\sqrt{1-w^2/c^2}$. So, calling the effective flow at such a moving position $U_w$, we have

$$U_w = U_0\sqrt{1-w^2/c^2} \qquad ((5\text{-}2)$$

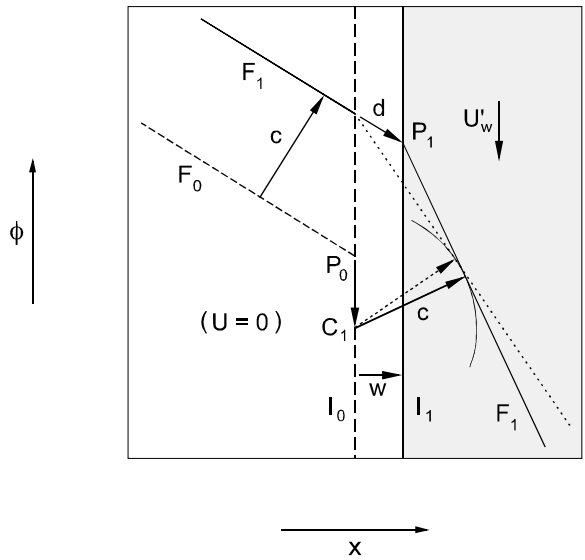

**Fig. 5-3.** Refraction by a moving ϕ flow region, with velocity w. $I_0$ and $I_1$ represent the interface at times t=0 and t=1 respectively, while $F_0$ and $F_1$ again are the refracting wavefront. (The refracted part of $F_0$ is omitted.) Here the interface retreats from the incident wavefront and the refraction effect is enhanced.

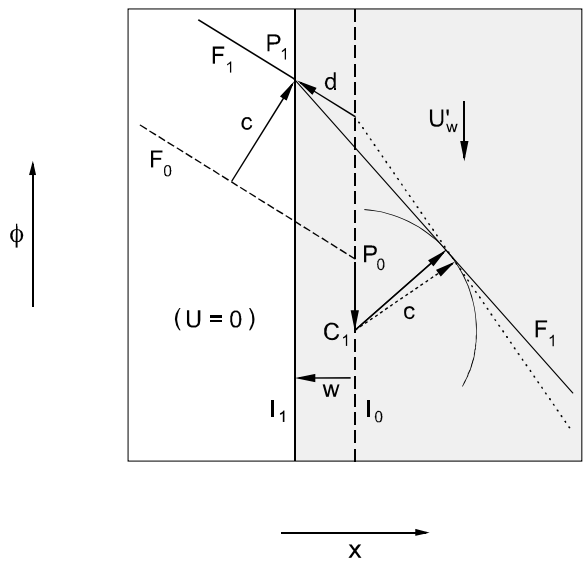

**Fig. 5-4.** Another case of refraction by a moving ϕ flow region. Instead of retreating, here the interface approaches the incident wavefront. In this case, the refraction effect is diminished.

We've seen that it's possible, in principle, to have wave-based, localized ϕ flows which move in x, y and z, without the displacement of the surrounding fluid in these dimensions. Given that our flows move in this fashion, we can derive the vector potential. Once more, we'll begin by looking at refraction at a sharp interface. Since this depends on the effective ϕ flow at a position moving with the interface, we'll also take it that the transformed ϕ flow is that given by Eq. (5-2), for an interface velocity, w.

For a qualitative introduction to the vector potential, we'll assume for the moment that our refracting ϕ flows are effectively steady. To visualize such flows, you could imagine that the responsible waves are of a much smaller scale than the waves being refracted, like the example of Fig. 5-1.

Refraction by effectively steady, moving ϕ flows is depicted in Figs. 5-3 and 5-4. Fig. 5-3 shows a ϕ flow retreating from an incident wavefront. Again, $F_0$ and $F_1$ represent a single refracting wavefront at times t = 0 and t = 1. (Only the incident portion of $F_0$ is shown.) $I_0$ and $I_1$ are the flow region interface at these times. The interface has a positive x velocity, indicated by the vector, w. Within the flow region, there is a downward flow, parallel to the ϕ axis, indicated by the vector $U_w'$. (Remember that the amplitude of $U_w'$ is also a function of the region's velocity, w.) The successive points of intersection between wavefront and interface are $P_0$ and $P_1$.

The semicircle in the figure represents a wavelet, originating from $P_0$, at time t = 0. Having been displaced by the flow $U_w'$, its center at time t = 1 is $C_1$. As before, by Huygens' principle, the refracted portion of wavefront $F_1$ is tangent to this wavelet and passes through $P_1$. However, unlike our earlier refraction examples, here $P_1$ is displaced by the interface movement. This displacement, indicated by the vector, d, is along the incident wavefront. As a result, in this case, the refraction is *increased* by the interface movement. For comparison, the refracted wavefront and pro-particle vector for a stationary interface are shown as a dotted line and arrow.

Fig. 5-4 shows the same incident wavefront, but with the flow interface moving toward it, in the negative x direction. As indicated by vector d, here $P_1$ is shifted oppositely along the incoming wavefront. In this case, the refraction effect is *diminished* by the interface movement.

# Wave Frequencies

Here we'll be looking at the effects of moving ϕ flows on the frequency of refracted waves in our system. They will be compared to the effects of moving scalar/vector potentials on de Broglie waves. Later, we'll use these wave frequencies to compare the velocity behaviors of pro-particles and actual charged particles.

Unlike our previous work with stationary ϕ flows, where we only treated the x dimension of the reference space, here we'll describe all three of its dimensions. What we're after is a general equation for the wave frequency, seen at an arbitrary moving point in the reference space. We'll call the 3-D velocity vector for this point **u**. Also, the waves of interest will be characterized by the 3-D velocity, **v**, of an associated pro-particle.

To begin, we'll write an expression for the frequency with no ϕ flow. In this case, the frequency of plane waves in the reference space* can be written as

$$\nu_{\mathbf{u}} = \nu - \left( \frac{\mathbf{u} \cdot \mathbf{v}}{v} \right) / \lambda \qquad (5\text{-}3)$$

Here ν is the frequency at a stationary position, $\nu_{\mathbf{u}}$ is that at an arbitrary moving point with velocity **u**, and λ is the wavelength. (Since **v**/v is a unit vector perpendicular to the wavefronts, the term in parentheses is the component of **u** in that direction.)

From Section 2, ν and λ are given in terms of the pro-particle velocity by

$$\nu = \frac{c}{s\sqrt{1 - v^2/c^2}} \qquad (5\text{-}4)$$

and

$$\lambda = \frac{sc\sqrt{1 - v^2/c^2}}{v} \qquad (5\text{-}5)$$

where s is the ϕ structure interval of the wave system. Putting these into Eq. (5-3) and simplifying gives

$$\nu_{\mathbf{u}} = \left( \frac{c}{s} \right) \frac{(1 - \mathbf{u} \cdot \mathbf{v}/c^2)}{\sqrt{1 - v^2/c^2}} \qquad (5\text{-}6)$$

where there is no ϕ flow.

---

*Our hyperplane waves appear as plane ones in the reference space.

Next, we'll write an equation for the effective flow of a moving ϕ flow region. Eq. (5-1) gives the effective flow at a stationary point in the reference space. Here we want this for an arbitrary moving point. As mentioned, the effective ϕ flow depends on the frequency of the underlying flow waves. Treating the ϕ flow as a second set of waves, we can use their frequency to find it.

The velocity, **w**, of a moving ϕ flow region corresponds to that of a pro-particle associated with its flow waves. Substituting **w** for the pro-particle velocity in Eq. (5-6), we can write this flow wave frequency as

$$\nu_u = \nu_0 \frac{(1 - \mathbf{u} \cdot \mathbf{w}/c^2)}{\sqrt{1 - w^2/c^2}} \tag{5-7}$$

where $\nu_0$ is the frequency of a stationary wave region. (Again, $\nu_0$ is equal to the quantity c/s.) To simplify the wave interactions, we assumed earlier that our flow waves have a relatively short wavelength. In this case, the values of s and $\nu_0$ for the flow waves are taken to be different than for those being refracted. Nevertheless, the transformation of this $\nu_0$ is the same.)

Since the effective flow transforms as the frequency does, we can also write $U_u$, the effective flow at a moving position, as

$$U_u = U_0 \frac{(1 - \mathbf{u} \cdot \mathbf{w}/c^2)}{\sqrt{1 - w^2/c^2}} \tag{5-8}$$

where $U_0$ is the corresponding ϕ flow at a stationary position, in a stationary flow region.

At a point with velocity **u**, the effect of such a flow on other waves is to shift the frequency by $QU_u/s$. (Again, Q has the value +1 for a positive pro-particle and -1 for a negative one.) From this and Eqs. (5-6) and (5-8), the frequency at an arbitrary moving position is then

$$\nu_u = \frac{1}{s} \left( c \frac{(1 - \mathbf{u} \cdot \mathbf{v}/c^2)}{\sqrt{1 - v^2/c^2}} + QU_0 \frac{(1 - \mathbf{u} \cdot \mathbf{w}/c^2)}{\sqrt{1 - w^2/c^2}} \right) \tag{5-9}$$

This is our general frequency equation.

Now we'll develop a similar equation for de Broglie waves in moving potential regions. Earlier, we looked at the effects of stationary scalar potential regions. Here we'll be concerned with scalar/vector potentials which are the moving, relativistic equivalent of these.

Where $\Phi_0$ is the stationary scalar potential, the relativistic transformation for a 4-vector gives

$$\Phi = \frac{\Phi_0}{\sqrt{1 - w^2/c^2}} \tag{5-10}$$

for the moving scalar potential, where **w** is the velocity of the potential region. In addition, there is now a vector potential*

$$\mathbf{A} = \frac{\mathbf{w}\,\Phi_0}{c\sqrt{1 - w^2/c^2}} \tag{5-11}$$

The vector potential is also given by the simple relationship

$$\mathbf{A} = \frac{\mathbf{w}}{c}\,\Phi \tag{5-12}$$

(This combination of scalar and vector potentials is that of individual charged particles. Of course *any* mixture of electromagnetic potentials can be attributed to the combined effects of such particles.)

*Note that, when we speak of scalar and vector potentials here, we are referring only to those in a single, specific reference frame. This will correspond to the preferred frame defined by the medium of our wave system.* (As Bell[15] reiterates, if a physical system behaves relativistically in only one frame, this is sufficient.)

As seen at a fixed spatial position, the frequency and wavelength of a particle's de Broglie waves are given by the Planck and de Broglie relations

$$\nu = E/h \tag{5-13}$$

and

$$\lambda = h/p \tag{5-14}$$

where h is Planck's constant, E is the particle's energy, and p its momentum.

---

*We'll use Gaussian units, where the vector potential has the dimensions of a potential.

To write a more general equation for the frequency at a moving point with velocity, **u**, we can again use Eq. (5-3), which we wrote initially for a pro-particle. Substituting the last two expressions into Eq. (5-3) and simplifying gives

$$\nu_u = \frac{1}{h}(E - \mathbf{u}\cdot\mathbf{p}) \tag{5-15}$$

Next we'll substitute for E and **p**. A familiar equation for the energy of a charged particle in both scalar and vector potentials is the Hamiltonian[19]

$$H = \sqrt{\left(\mathbf{p} - \frac{q}{c}\mathbf{A}\right)^2 + m^2 c^4} + q\Phi \tag{5-16}$$

However, the relativistic momentum, **p**, is given by

$$\mathbf{p} = \frac{m_0 \mathbf{v}}{\sqrt{1 - v^2/c^2}} + \frac{q}{c}\mathbf{A} \tag{5-17}$$

Rewriting H as E, substituting the above expression for **p**, and also putting the particle mass, m, in terms of $m_0$ and v, Eq. (5-16) reduces to

$$E = \frac{m_0 c^2}{\sqrt{1 - v^2/c^2}} + q\Phi \tag{5-18}$$

which is just the energy equation used in the last section. So, for a given particle velocity, the energy is actually independent of **A**.

Putting these expressions for E and **p** into Eq. (5-15) and rearranging gives

$$\nu_u = \frac{1}{h}\left(m_0 c^2 \frac{(1 - \mathbf{u}\cdot\mathbf{v}/c^2)}{\sqrt{1 - v^2/c^2}} + q\Phi - \frac{q\mathbf{u}\cdot\mathbf{A}}{c}\right) \tag{5-19}$$

Finally, from Eqs. (5-10) and (5-11), we can substitute for $\Phi$ and **A** to obtain

$$\nu_u = \frac{1}{h}\left(m_0 c^2 \frac{(1 - \mathbf{u}\cdot\mathbf{v}/c^2)}{\sqrt{1 - v^2/c^2}} + q\Phi_0 \frac{(1 - \mathbf{u}\cdot\mathbf{w}/c^2)}{\sqrt{1 - w^2/c^2}}\right) \tag{5-20}$$

*Here again, we see a precise analogy between the wave frequencies associated with pro-particles and charged particles.* With our earlier substitutions of $h/m_0 c$ for s, and $q\Phi/m_0 c$ for QU, Eq. (5-9) for a pro-particle gives becomes this one.

## Pro-particle and Particle Velocities

Earlier, we described the velocity of a charged particle entering a stationary scalar potential region. To solve for the velocity, we used the conservation of energy for a "free" particle. However, the energy of a relativistic particle entering a *moving* potential region is no longer conserved.* Instead, here we'll use a conserved de Broglie wave frequency. For a potential interface in motion, at its moving position, the wave frequencies on both sides must match. Then we'll compare the velocity equations obtained to those for to pro-particles entering moving ϕ flows

For an actual particle, we'll begin by writing

$$\nu_{\mathbf{w}}' = \nu_{\mathbf{w}} \qquad (5\text{-}21)$$

where these are the de Broglie frequencies on both sides, at the interface of a scalar/vector potential region with velocity **w**. Again, primes refer to quantities inside the potential.

From Eq. (5-20), setting the position vector, **u**, equal to the potential region's velocity, **w**, we can write expressions for the two sides of Eq. (5-21). Outside the potential region, this gives

$$\nu_{\mathbf{w}} = \frac{1}{h}\left(m_0 c^2 \frac{(1-\mathbf{v}\cdot\mathbf{w}/c^2)}{\sqrt{1-v^2/c^2}}\right) \qquad (5\text{-}22)$$

In the potential region, we have

$$\nu_{\mathbf{w}}' = \frac{1}{h}\left(m_0 c^2 \frac{(1-\mathbf{v}'\cdot\mathbf{w}/c^2)}{\sqrt{1-v'^2/c^2}} + q\Phi_0'\sqrt{1-w^2/c^2}\right) \qquad (5\text{-}23)$$

Putting these expressions into Eq. (5-21) and multiplying both sides by $h/m_0 c$ gives this governing equation for the particle velocity:

$$\frac{c - \mathbf{v}'\cdot\mathbf{w}/c}{\sqrt{1-v'^2/c^2}} + \frac{q\Phi_0'}{m_0 c}\sqrt{1-w^2/c^2} = \frac{c - \mathbf{v}\cdot\mathbf{w}/c}{\sqrt{1-v^2/c^2}} \qquad (5\text{-}24)$$

To illustrate how this translates into particle velocity, we'll take the case of one-dimensional motion, where **v** and **w** are parallel and can be treated as scalars. It can be verified that the

---

*Although not really bound, such a particle is not called "free".

solution for v' in this case is given by

$$v' = c \left( \frac{\pm K \sqrt{K^2 + w^2 - c^2} + wc}{K^2 + w^2} \right) \qquad (5\text{-}25)$$

where the sign of the square root corresponds to that of v, and K is defined by

$$K = \frac{c - vw/c}{\sqrt{1 - v^2/c^2}} - \frac{q\Phi_0'}{m_0 c}\sqrt{1 - w^2/c^2} \qquad (5\text{-}26)$$

(When w is zero, these equations reduce to Eq. (4-13) for the particle velocity in a stationary scalar potential.)

For a pro-particle, starting with Eqs. (5-21) and (5-9) and following the same basic procedure leads to an analogous governing equation for the velocity:

$$\frac{c - \mathbf{v}' \cdot \mathbf{w}/c}{\sqrt{1 - v'^2/c^2}} + QU_0'\sqrt{1 - w^2/c^2} = \frac{c - \mathbf{v} \cdot \mathbf{w}/c}{\sqrt{1 - v^2/c^2}} \qquad (5\text{-}27)$$

*Hence frequency translates into velocity in exactly the same way for both pro-particles and particles.*

For the case of one-dimensional motion in the reference space, where **v** and **w** are parallel, solving for v' gives Eq. (5-25) again, where K is now defined as

$$K = \frac{c - vw/c}{\sqrt{1 - v^2/c^2}} - qU_0'\sqrt{1 - w^2/c^2} \qquad (5\text{-}28)$$

(The solution for Eq. (5-27) was found geometrically from the wave representation provided by this theory. While the derivation is omitted for brevity, I think its directness illustrates the utility of the 4-D Euclidean approach. The solution given previously for a real particle was actually gotten from this one.)

Eq. (5-20) for the de Broglie wave frequency is an interesting and versatile one. By setting the position vector, **u**, equal to the potential region velocity, **w**, we obtained our governing equation for particle velocity. We can also set **u** equal to 0, for the frequency at a stationary position. From E = hv, multiplying this frequency by h gives Eq. (5-18), the Hamiltonian of the particle. In addition, the Lagrangian can be obtained similarly. This is related to the frequency at the *particle's* moving position. The usual form of the relativistic Lagrangian is[19]

$$L = -m_0 c^2 \sqrt{1 - v^2/c^2} - q\Phi + q\mathbf{A} \cdot \mathbf{v}/c \qquad (5\text{-}29)$$

For a potential described by Eqs. (5-10) and (5-11), this can also be written as

$$L = -m_0 c^2 \sqrt{1 - v^2/c^2} - \frac{q\Phi_0 (1 - \mathbf{v} \cdot \mathbf{w}/c^2)}{\sqrt{1 - w^2/c^2}} \qquad (5\text{-}30)$$

Setting **u** equal to **v** in Eq. (5-20) yields the de Broglie frequency at the particle's position. Simplifying and multiplying by -h then gives the Lagrangian above. (The reason for a negative h here is that the potential energy component of the Lagrangian is defined as negative, while this is positive in the Hamiltonian.)

In classical electrodynamics, the Lagrangian gives rise to the force equation for a charged particle in electric and magnetic fields. (For example, see Duffy[20].) In Gaussian units, this is

$$\mathbf{F} = q\left(\mathbf{E} + \frac{\mathbf{v} \times \mathbf{B}}{c}\right) \qquad (5\text{-}31)$$

where **F** is the relativistic force defined by

$$\mathbf{F} = \frac{d}{dt}\left(\frac{m_0 \mathbf{v}}{\sqrt{1 - v^2/c^2}}\right) \qquad (5\text{-}32)$$

**E** is the electric field given by

$$\mathbf{E} = -\nabla \Phi - \frac{1}{c}\frac{\partial \mathbf{A}}{\partial t} \qquad (5\text{-}33)$$

and **B** is the magnetic field from

$$\mathbf{B} = \nabla \times \mathbf{A} \qquad (5\text{-}34)$$

Similarly, a quantity which behaves like the Lagrangian of Eq. (5-30) can be obtained for a pro-particle, from the frequency at its moving position. For example, setting **u** equal to **v** in Eq. (5-9), and multiplying by -s gives an analogous quantity:

$$-c\sqrt{1 - v^2/c^2} - QU_0 \frac{(1 - \mathbf{v} \cdot \mathbf{w}/c^2)}{\sqrt{1 - w^2/c^2}} \qquad (5\text{-}35)$$

From this and the usual derivation for Eq. (5-31) from the Lagrangian, it would be possible to write an analogous "force" equation for pro-particles. This would give their behavior in continuous φ flow gradients, as opposed to that across sharp interfaces. However, because Eq. (5-31) (which contains **E** and **B**) is only valid in classical electrodynamics, we won't pursue its analogue here.

As described in Feynman, Leighton and Sands,[16] electric and magnetic fields don't account for quantum mechanical phenomena, such as the Aharanov-Bohm effect. At the quantum level, electromagnetic effects are seen instead in terms of the influence of scalar and vector potentials on the phases of de Broglie waves. This can correctly be expressed as

$$\Delta phase = \frac{q}{h}\int \Phi dt - \frac{q}{hc}\int \mathbf{A}\cdot d\mathbf{s} \qquad (5\text{-}36)$$

where Δphase is measured in cycles, the latter integral is taken over a possible particle trajectory, and d**s** is an element of the trajectory. As Feynman points out, these two integrals are the quantum mechanical replacement for Eq. (5-31) for the force on a charged particle.

Substituting **v**dt for d**s**, this can be rewritten as

$$\Delta phase = \frac{q}{h}\int (\Phi - \mathbf{v}\cdot \mathbf{A}/c)dt \qquad (5\text{-}37)$$

For potentials associated with uniformly moving charges, from Eq. (5-12), this becomes

$$\Delta phase = \frac{q}{h}\int \Phi(1 - \mathbf{v}\cdot \mathbf{w}/c^2)dt \qquad (5\text{-}38)$$

*Our 4-D Euclidean system provides a simple physical basis for such phase shifts: the displacement of waves in the ϕ dimension.* We have

$$\Delta phase = Q\frac{\Delta \phi}{s} \qquad (5\text{-}39)$$

where Δϕ represents the ϕ displacement and s is the ϕ period of the waves. Of course Δϕ corresponds to the effective ϕ flow, integrated over time. From the effective flow given by Eq. (5-8), put in terms of U instead of $U_0$ from Eq. (5-1), we can also write

$$\Delta phase = \frac{Q}{s}\int U(1 - \mathbf{v}\cdot \mathbf{w}/c^2)dt \qquad (5\text{-}40)$$

This describes the analogous case of a pro-particle in a uniformly moving ϕ flow region. (Not necessarily with sharp boundaries.) The vector potential is represented in Eq. (5-38) by the quantity Φ**w**/c. We have a corresponding quantity, U**w**/c, in the last equation, as part of the term -U**v**·**w**/$c^2$. Again, this term represents a change in the effective ϕ flow at the moving position of a pro-particle with velocity **v**, as more or fewer flow waves are encountered. *This effect is the basis for the vector potential in this theory.*

In this section, we've carefully avoided the issue of mutual flow wave interactions. Here we've only looked at the behavior of small-amplitude waves, in ϕ flows driven by relatively short-wavelength flow waves. In this artificial case, there is no reciprocal effect of the former waves on the latter. (This is basically the same assumption made in the study of electromagnetism, when test charges are taken to have no effect on surrounding fields.)

Still, the flow waves in this system *do* interact mutually. In this case, feedback arises. For example, wave set A refracts and reflects set B, which, by altering B's effect on A, then changes A -- and so forth. While such effects can be modeled by computer simulation, they are difficult to treat analytically in more than one dimension.

Typically, feedback in nonlinear wave systems leads to pattern formation. While the waves in this system remain repetitive in the ϕ dimension, the nonlinear nature of flow waves should lead to additional patterns within the overall structure. Such self-organizing patterns are taken as the basis for the realistic particle model introduced in the next section.

# 6. Wavicles

Just as this theory involves a nonstandard interpretation of relativity, an alternate interpretation of quantum mechanics is also involved. As we'll see below, in the context of this wave system, this permits a new, realistic model of elementary particles.

In the usual textbook interpretation, de Broglie waves are taken to be linear waves of pure probability, whose squared moduli determine the likelihood of finding particles in particular states. These nonphysical waves propagate through increasing volumes of space (at speeds greater than light) until instantaneously "reduced" by a measurement. In the times between measurements, particles are taken not to exist in any definite state.

However, as discussed by Bell[21], what constitutes a "measurement" has never been clearly defined. Despite its widespread acceptance and its usefulness as a calculational tool, the standard interpretation of quantum mechanics brings severe conceptual difficulties. This has also been pointed out by de Broglie[1], Schrödinger[22], Einstein[23], and Dirac[24] -- each an architect of the existing quantum theory.

A second, distinctly different interpretation of quantum mechanics exists, which is equally consistent with experiment. (At least in the nonrelativistic case.) This is the "hidden" variables theory of de Broglie, Bohm, and Vigier. Here, de Broglie waves are taken to be real, nonlinear ones, and particles have definite states in the times between measurements.* In such a theory, Heisenberg uncertainty derives from inherent limitations in the measurement process. The present theory can be considered a relativistic extension of this one.

Until Bohm's paper of 1952[2], it was generally held that a self-consistent hidden variables theory was impossible. A frequently cited proof to this effect had been published by von Neumann in 1932. (Although a serious flaw in von Neumann's proof was pointed out by Grete Hermann in 1935, her finding went largely unnoticed at the time.) Bohm refuted this idea by actually constructing a self-consistent theory of this type. As Bell put it later, "in 1952, I saw the impossible done".[26]

Bell's Theorem was developed in response to Bohm's work, to define the requirements for a valid hidden variables theory. It indicates that, in *any* self-consistent theory of quantum mechanics, there are necessarily non-local effects, as there are in Bohm's. While Bell was a proponent of hidden variables, their required non-local character has been taken by some as evidence against their existence. However, it will be shown that similar non-local phenomena can be found in ordinary macroscopic systems. So, while the present theory will be seen to involve particle-like entities, existing more-or-less continuously in definite states, this is not in conflict with the usual laws of quantum mechanics.

---

*Albert, Aharanov and D'Amato[25] have also given strong theoretical evidence that particles *must* have definite states between measurements.

As mentioned, here elementary particles and their fields are based on self-organizing wave patterns. What types of patterns might be expected to form within this repetitive, nonlinear system? In the absence of computer simulation, we'll take as a guide the behavior of nonlinear, three-dimensional acoustic waves in ultrasonic wave tanks.

A characteristic effect arising in ultrasonic tanks is acoustic cavitation. This involves the formation of microscopic bubbles, which expand and collapse with impinging waves.* While bubbles are easily formed when the liquid contains dissolved gasses, these also arise in purified, degassed liquids. Here they consist of a vapor or other form of the primary liquid. In clouds of such bubbles, individual ones can exhibit surprising persistence. A variety of theories have been proposed to explain this, many involving assumed impurities in the liquid.[28]

Usually ignored in such theories is the self-organizational capacity of strongly nonlinear wave systems. While waves determine where the bubbles are**, the latter also influence the propagation of waves (directly by reflection and also indirectly through the nonlinear interaction of reflected and other waves). So it may be that these pulsating bubbles are sustained by feedback, where waves are directed *toward* the bubble sites. We'll assume that this is so. (This would also help explain the large wave amplifications observed.) Bubbles then occupy the centers of more-or-less spherical standing wave patterns, with amplitudes varying as the reciprocal square of the distance to the bubble center.

Could similar patterns arise in our 4-D wave system? The simplest 4-D analogy would involve hyperspherical waves. However, in this repetitive system, isolated hyperspherical patterns can't exist. Instead, there would necessarily be an overlapping series of these, repeating in the ϕ dimension with the structure interval, s. At some distance from the x, y, z center, the fronts of the combined waves would approximate hypercylinders, aligned with the ϕ axis. Like 3-D spherical ones, the average amplitudes for such waves would vary as $1/r^2$, where the radius, r, parallels the reference space. (For isolated hyperspherical waves, it would be $1/r^3$.)

Notice that, while we are constrained to repeating waveforms, our four-dimensional system still has another degree of freedom not found in a three-dimensional one. Besides hypercylindrical patterns, hyper*conical* ones of various shapes are also permitted. (The former are a special case of these.) Fig. 6-1 depicts sections of several hyperconical fronts (resembling nested paper cups), with respect to ϕ, plus an arbitrary two of the three reference space dimensions. These are arrayed along a common axis paralleling the ϕ dimension, and again repeat with interval s.

---

*A remarkable aspect is the extreme concentration of wave energy occurring during bubble collapse. In water, this can cause a bluish or violet light emission called sonoluminescence. Energy gains of over $10^{12}$ have been reported. The energies attained are so great that, for bubbles containing a deuterium/tritium mixture, measurable thermonuclear fusion may be possible.[27]

**This is demonstrated clearly in single-bubble sonoluminescence experiments, where the bubble's position is governed by focussed waves, overriding its gravitational tendency to rise.[29]

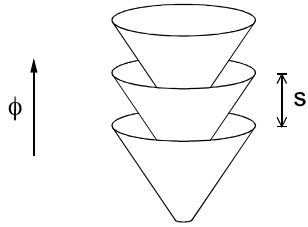

**Fig. 6-1.** Hyperconical wavefront sections, depicted in terms of ϕ, plus an arbitrary two of the three reference space dimensions. These represent a wave array which extends indefinitely in all directions, repeating in ϕ with the usual structure interval, s.

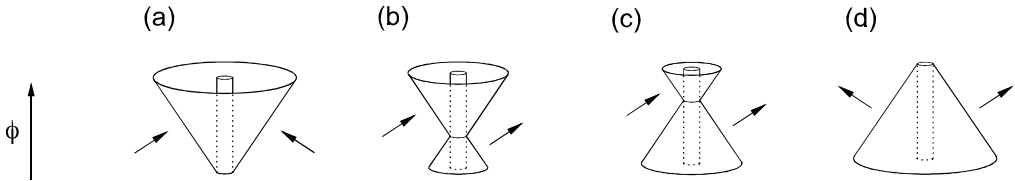

**Fig. 6-2.** For a simplified case, (a)-(d) illustrate the progressive reflection of a hyperconical wavefront section by a nonlinear "core" region.

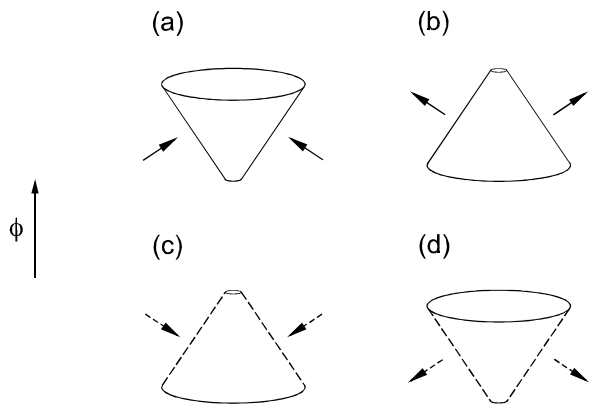

**Fig. 6-3.** These wavefronts characterize four wave arrays comprising a single "wavicle." As indicated by solid lines, (a) and (b) represent incident and reflected waves of positive density. In (c) and (d), broken lines signify corresponding waves of negative relative density, moving oppositely with respect to ϕ. A $1/r^2$ ϕ flow is associated with the waves.

Within a central "core" region, where the amplitudes are great, we'll consider these hyperconical waves to be nonlinear. Consequently, they are reflected and/or transmitted through this region with phase shifts. For simplicity, here we'll take it that the reflection is total and abrupt, as shown in Fig. 6-2. As indicated by arrows, the cone in (a) represents an incident wavefront, while the thin cylinder at the center indicates the nonlinear core. The front's progressive reflection is depicted in (a)-(d), such that in (d) the wavefront is entirely reflected.

Assuming that the waves in these patterns are flow waves, then, from the last section, for every wave of positive relative density we can also expect a balancing negative one. Combined in the same pattern, this would give four wave sets all together, as illustrated in Fig. 6-3. While (a) represents an incident wave of positive density, (b) represents a corresponding wave of like density after reflection. Both move positively with respect to ϕ. As indicated by broken lines, (c) and (d) are the corresponding incident and reflected waves of negative density, moving negatively in the ϕ dimension. Of course each of these represents an array of hyperconical waves, with these crisscrossing wave sets sharing the same core region.

*Associated with such flow wave patterns are net ϕ flows, corresponding to the ϕ components of the wave vectors.* (There is no net flow with respect to the reference space.) *These flows are proportional to the wave amplitudes, which, again, vary as $1/r^2$.* So a wavicle contains a stream, whose flow increases towards the core. Inverted patterns, with opposite ϕ flows (charges), are also permitted.

A key assumption in this theory is that, while the system is driven to form such patterns, *the ϕ flow attainable at the cores has a sharp limit*. This limiting velocity is regarded as a general characteristic of the wave system, like the characteristic wave velocity, c. With positive feedback acting to increase the core flow, this limit then becomes the defining nonlinear element of these patterns, setting the flow wave amplitudes.

Attributing electromagnetic potentials to ϕ flows, as described earlier, we then have particle-like entities with quantized charge, energy (mass), and proper Coulomb fields. Also, particle "spin" can be attributed to an extra degree of freedom of these 4-D these standing wave patterns. In an introduction to quantum electrodynamics, Richard Feynman remarked that, since they also behave as waves, elementary particles could just as well be referred to as "wavicles." Here we'll adopt Feynman's term as the name for these hyperconical flow wave patterns, which form the basis for charged particles in this theory.

(The axially symmetric configuration of Figs. 6-1 to 6-3 only characterizes a stationary wavicle. As described below, at relativistic speeds, the hyperconical fronts are tipped with respect to the core, and open out. Also, in an actual wavicle, the wavefronts would be somewhat irregular, unlike the perfectly symmetrical hypercones depicted.)

Of course wavicles depend on incoming waves, originating in remote parts of the system, for their continued existence. Thus, *unlike the usual conception of a particle, they can't be considered to exist independently of their surrounding environment*. The source of a wavicle's incoming waves is taken to be others in the system. Likewise, its outgoing waves are also incoming ones for those. (Like the reflecting waves in a large cloud of cavitation bubbles.) Consequently, *the waves comprising a wavicle are not exclusively its own*. (This is manifested in the inseparability of particles and observing apparatus in quantum-mechanical experiments.)

For a conventional charged particle to have finite size, additional "Poincaré stresses" must be introduced to hold it together against the effects of its own field.[30] (Finite size avoids the serious problem of an infinite field at the particle's location.) One immediate advantage of wavicles is that no Poincaré stresses are called for, since these are contained naturally by the inward momentum of their incoming waves. (The wavicle model also seems to answer the long-standing problem of the basis of radiative reaction in charged particles. This calls for "half-advanced" and "half-retarded" waves, where the former anticipate the future particle position.[31])

## Moiré Wavefronts

Another premise of this theory is that certain wavicle patterns are more stable than others. These have characteristic wave angles and frequencies, in the stationary case, and correspond to different elementary particles or sub-particles. As proposed by de Broglie, we'll take the frequency and mass to be proportional, via the Planck and Einstein relations

$$h\nu = E = mc^2 \qquad (6\text{-}1)$$

Given a stationary wavicle, a moving one is obtained by applying our usual relativistic transformation, shown to hold for this system. (The defining conditions for a wavicle are taken to transform along with its waves.) While the component waves have a single frequency in the rest case, this is no longer true when a wavicle moves. However, in combination, they comprise "Moiré wavefronts," which still have a single frequency. Here we'll find that such wavefronts correspond, in more than one way, to the de Broglie waves of particles with various rest masses.

Fig. 6-4 illustrates a cross-section of a stationary wavicle in two dimensions, showing only positive density waves. The two crisscrossing sets, marked by thin lines, are the same waves, before and after reflection from the core. (The core is shown as the vertical gap at the center.) In this cross-section, these waves have just two orientations, with incident and reflected waves on opposite sides of the core sharing the same angle. As denoted by vectors with magnitude c, the waves of both orientations are moving positively with respect to $\phi$. Together, they form an unchanging pattern, moving parallel to the $\phi$ axis. The motion of the pattern's nodes is indicated on the right by the vector, N.

**Fig. 6-4.** Cross-sectional view of a stationary wavicle, in x and φ. The central gap represents the core, while thin lines again represent component waves, with velocities, c. Vector N indicates the motion of nodes and the wave pattern as a whole. The pattern's overall phases are characterized in terms of "Moiré wavefronts," depicted here as horizontal gray bars.

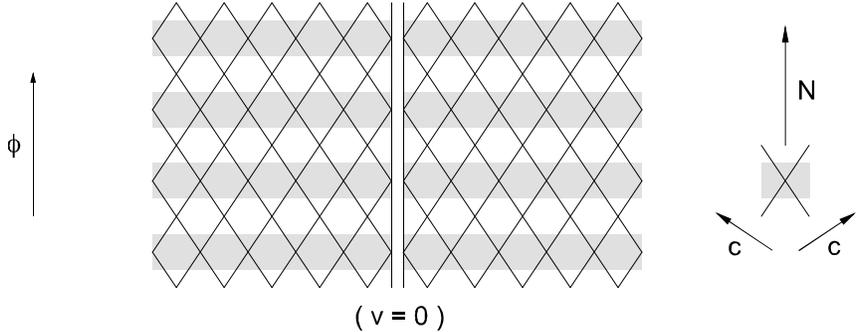

( v = 0 )

**Fig. 6-5.** This shows the same wavicle, moving at .3c in the +x direction. Part (b) expands an element of the wave pattern, with a Moiré wavefront represented here by a dotted line, midway between wavefronts a and b. (The vertical segment s is an added construct.) Distance d/2 is Lorentz contracted.

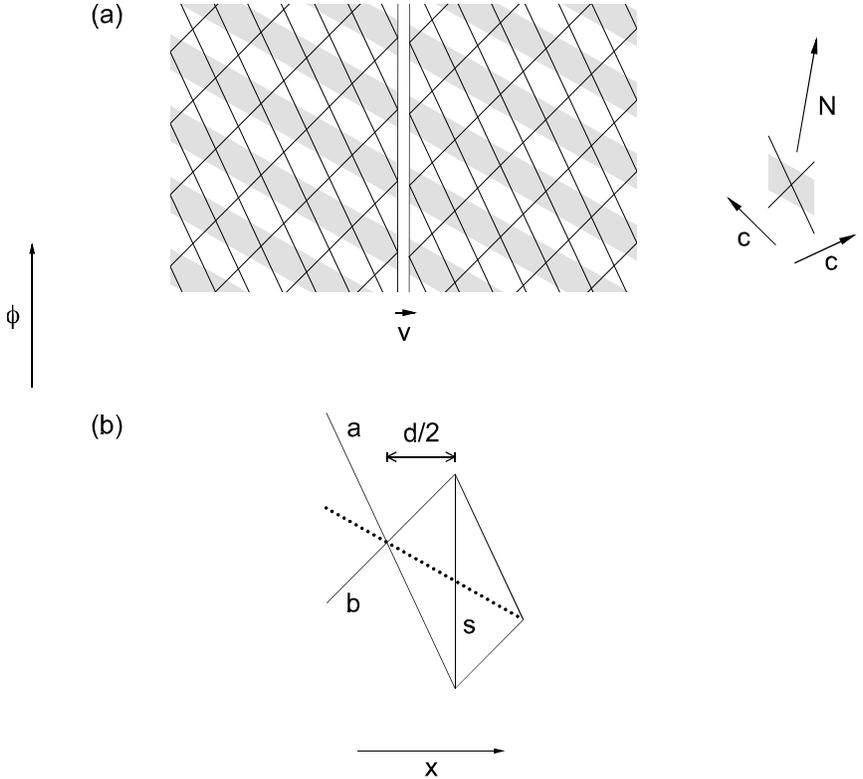

A striking feature of any standing wave pattern is that, *regardless of its size, all regions undergo synchronized behavior*. As seen in the reference space, the extended wave field of a stationary wavicle undulates or pulsates in unison, at a specific frequency. This corresponds to the φ movements of the component wave patterns through the reference space. It's these pattern movements that we'll be describing in terms of Moiré wavefronts.

If the positive component waves of Fig. 6-4 were drawn with thick lines, from the Moiré effect, you might see alternating light and dark bands, running horizontally across the nodes. Moiré wavefronts, depicted here by gray bars, have the same orientation; hence the term. *Representing phases of the overall wave pattern, Moiré wavefronts are defined as having the same φ period, s.* (This omits half the lines which would be perceived visually with a Moiré pattern.) Thus, in the reference space, *their frequency represents that of a wavicle's pulsations*.

For a stationary wavicle, the component wave and Moiré frequencies are all identical. As before, we'll characterize the component waves in terms of associated pro-particle velocities. (Knowing the relativistic transformation equations for pro-particles, we can then use these to describe the transformation of wavicles.) Working in just the x and φ dimensions of Fig. 6-4, with representative positive density waves of just two orientations, we'll call the characteristic pro-particle velocities $u_a'$ and $u_b'$. So we have

$$u_a' = -u_b' \qquad (6\text{-}2)$$

The wave frequency corresponding to a given pro-particle velocity is given by Eq. (2-7). Inserting $u_a'$ and $u_b'$ into that equation, we can write

$$\nu_0 = \frac{c}{s\sqrt{1 - u_a'^2/c^2}} = \frac{c}{s\sqrt{1 - u_b'^2/c^2}} \qquad (6\text{-}3)$$

with $\nu_0$ referring to the "rest" Moiré frequency. Again, greater pro-particle velocities would represent more tilted component waves, and a higher rest frequency. From Eq. (6-1), this frequency is taken to correspond to a particle's rest mass, by the relationship

$$\nu_0 = m_0 c^2 / h \qquad (6\text{-}4)$$

Fig 6-5 depicts the same wavicle, moving at a relativistic speed, v. In addition to the core, the pattern nodes must also have this x velocity. (Since the component wavefronts are parallel within each set and meet at nodes at the core surface.) Again, the component waves have just two orientations, with incident and reflected waves on opposite sides of the core matching in angle. As shown, the Moiré wavefronts are tilted in this case.

For a moving wavicle, at a position traveling with it, we know the frequencies of both component wave sets are decreased by simple relativistic time dilation. (This can also be shown formally from the transformation equations.) At such a position (the core, for example) the component wave and Moiré frequencies again are identical and can be expressed as

$$\nu_v = \nu_0 \sqrt{1 - v^2/c^2} \qquad (6-5)$$

where $\nu_v$ is that at a moving x position having the wavicle's velocity, v.

As noted, at a fixed position in the reference space, the component frequencies of a moving wavicle no longer match. In this case, the Moiré frequency corresponds to their *average*. This can be seen from Fig. 6-5(b), which expands an element of the wave pattern. Added to this parallelogram is a vertical diagonal segment, s (labeled by its length), separating wavefronts a and b in the ϕ dimension. Since the Moiré front bisects this, its ϕ coordinates must be the average of a and b's. It follows that the slope, phase velocities, and frequency of the Moiré wavefronts are also the averages of these quantities for the component waves.

Thus, from the frequency expressions for the component waves, the fixed-position Moiré frequency of a moving wavicle can be written in terms of the component frequencies as

$$\nu = \frac{c}{2s} \left( \frac{1}{\sqrt{1 - u_a^2/c^2}} + \frac{1}{\sqrt{1 - u_b^2/c^2}} \right) \qquad (6-6)$$

where $u_a$ and $u_b$ are the characteristic pro-particle velocities for the component waves. Since pro-particles have been shown to behave relativistically, $u_a$ and the characteristic velocity, $u_a'$, are related by the addition of velocities equation

$$u_a = \frac{u_a' + v}{1 + u_a' v/c^2} \qquad (6-7)$$

where v is the velocity of a reference frame moving with the wavicle. From Eq. (6-2), $u_b$ can also be written in terms of $u_a'$ as

$$u_b = \frac{-u_a' + v}{1 - u_a' v/c^2} \qquad (6-8)$$

Inserting these expressions for $u_a$ and $u_b$, it can be shown that Eq. (6-6) reduces to this one:

$$\nu = \frac{\nu_0}{\sqrt{1 - v^2/c^2}} \qquad (6-9)$$

where the rest frequency, $\nu_0$, is that given by Eq. (6-3). Hence this frequency *increases* relativistically.

The moving and fixed position frequencies are related by

$$\nu_v = \nu - v/\lambda \tag{6-10}$$

where $\lambda$ is the Moiré wavelength in the reference space. For $\lambda$, this gives

$$\lambda = \frac{v}{\nu - \nu_v} \tag{6-11}$$

Substituting for $\nu$ and $\nu_v$ from Eqs. (6-5) and (6-9), and then multiplying top and bottom by the square root underneath, we can also write

$$\lambda = \frac{v}{\nu_0/\sqrt{1-v^2/c^2} - \nu_0\sqrt{1-v^2/c^2}}$$
$$= \frac{c^2\sqrt{1-v^2/c^2}}{\nu_0 v} \tag{6-12}$$

Multiplying this wavelength by the frequency of Eq. (6-9) gives the Moiré phase velocity seen in the reference space,

$$V = c^2/v \tag{6-13}$$

which is independent of $\nu_0$. *Once again, we obtain the phase velocity of de Broglie waves.*

Substituting for $\nu_0$ from Eq. (6-4), Eqs. (6-9) and (6-12) become

$$\nu = \frac{m_0 c^2}{h\sqrt{1-v^2/c^2}} = \frac{E}{h} \tag{6-14}$$

and

$$\lambda = \frac{h\sqrt{1-v^2/c^2}}{m_0 v} = \frac{h}{p} \tag{6-15}$$

*the de Broglie frequency and wavelength for an actual particle, now of arbitrary rest mass, $m_0$.* In his original thesis, de Broglie hypothesized that particles have some "periodic inner process" corresponding to their masses, with which waves are associated. Here we find an appropriate process, in the inherent pulsations of wavicles. *We also have a physical basis for the relativistic rest energy of elementary particles, and the equivalence of mass and energy.*

(Here we've seen that Moiré wavefronts inherit the basic relativistic behavior of their component waves. However, in four-dimensional terms, there are significant differences. One obvious difference is ϕ velocity. This is greater for a Moiré front, and contributes to a 4-D velocity greater than c. Comparing the transformed wavefronts of stationary pro-particles and wavicles, the latter's Moiré fronts have a greater slope. Thus, as a function of distance in the reference space, these manifest larger ϕ offsets. Nevertheless, from its greater ϕ velocity, a Moiré front still arrives at the reference space at the proper relativistic time.)

Next, in terms of its node spacing, we'll describe the contraction of a moving wavicle, having velocity v in the x dimension. We'll call the x distance between successive nodes on the same Moiré wavefront d. Referring to Fig. 6-5(b) again, from the node where wavefronts a and b cross, to the segment, s, (of this ϕ length) the distance is d/2. From the figure, this distance, the slopes of a and b, and s are related by

$$\frac{d}{2} slope_b - \frac{d}{2} slope_a = s \qquad (6\text{-}16)$$

For d, we have

$$d = \frac{2s}{slope_b - slope_a} \qquad (6\text{-}17)$$

The component wave slopes are related to their characteristic pro-particle velocities by

$$slope = \frac{-u}{\sqrt{c^2 - u^2}} \qquad (6\text{-}18)$$

(See Fig. 2-3, for example.) To describe a stationary wavicle, we'll insert the pro-particle velocities of Eq. (6-2) into this expression. Then putting these slopes into Eq. (6-17), we get

$$d_0 = \frac{2s}{u_a'/\sqrt{c^2 - u_a'^2} + u_b'/\sqrt{c^2 - u_b'^2}} \qquad (6\text{-}19)$$

with $d_0$ the distance in the rest case. Similarly, we can write the general equation

$$d = \frac{2s}{u_a/\sqrt{c^2 - u_a^2} + u_b/\sqrt{c^2 - u_b^2}} \qquad (6\text{-}20)$$

with $u_a$ and $u_b$ representing the characteristic velocities of a moving wavicle. Putting this in terms of $u_a'$ and $u_b'$, using Eqs. (6-7) and (6-8), the result can be shown to be equivalent to the

expression

$$d = d_0\sqrt{1 - v^2/c^2} \tag{6-21}$$

where $d_0$ is given by Eq. (6-19). *So we again have Lorentz/Fitzgerald contraction. Since the wavicle core is defined by nodes, this is also contracted, in addition to the overall wave pattern.*

(Here I'd like to mention another difference between the transformations of component waves and Moiré wavefronts. Recall that, as the former are tilted in four dimensions, the spacing between associated pro-particles is invariant. However, because comparable Moiré fronts undergo greater tilts, the 4-D spacing of the associated nodes *increases* with tilt. In terms of the reference space, though, the node spacings undergo the same relativistic contraction as pro-particles.)

### 4-D Wavefront Transformations

Fig. 6-6 shows how individual hyperconical wavefronts transform for a wavicle with velocity v in the +x direction. Part (a) illustrates the configuration in $\phi$ and x, plus one of the remaining dimensions. As shown, the hyperconical axes remain parallel to the node velocity vector, N, which is tipped here with respect to the core and $\phi$ axis. Part (b) depicts the relationship between N, v, and vector $V_{\phi v}$, which is the $\phi$ phase velocity of a Moiré wavefront at an x position moving with velocity v. (For example, consider the $\phi$ motion of a node at the moving core surface.) From v and $V_{\phi v}$, we can find N.

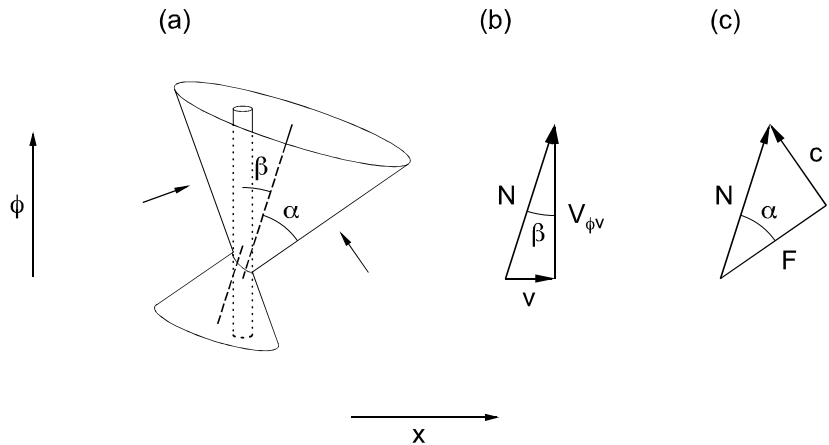

**Fig. 6-6.** Component wavefronts, from the previous wavicle, for a velocity of .5c in the +x direction. This depicts the configuration in $\phi$, x, and one remaining dimension. As illustrated, the node velocity vector, N, shown in (b) and (c) parallels the hypercone axes. Part (b) gives the angular tilt, $\beta$, of the axes with respect to the core. The angle $\alpha$ between the axes and hypercone surfaces is given by (c).

We can get an expression for $V_{\phi v}$ from the moving position Moiré frequency, $v_v$, provided by Eq. (6-5). Eliminating $v_0$ via Eq. (6-3) and multiplying by the $\phi$ structure interval, s, gives

$$V_{\phi v} = \sqrt{\frac{c^2 - v^2}{c^2 - u'^2}} \qquad (6\text{-}22)$$

where u' is the characteristic pro-particle velocity for the wavicle at rest. From the figure, putting N in terms of v and $V_{\phi v}$, we have

$$N = \sqrt{v^2 + V_{\phi v}^2} = \sqrt{v^2 + \frac{c^2 - v^2}{c^2 - u'^2}} \qquad (6\text{-}23)$$

(In analyzing this system, a possible source of confusion is the fact that the 4-D Moiré wavefront velocity is somewhat less than N. This results from the fact that, in a moving wavicle, N and the Moiré wavefronts are not exactly perpendicular.)

From the figure, the angle $\beta$, between N and the $\phi$ axis is given by

$$\sin\beta = v/N \qquad (6\text{-}24)$$

From Eq. (6-23), notice that as the wavicle velocity, v, approaches c, N also approaches c, as a lower limit. Thus for values of v close to c, the ratio on the right in the last equation approaches 1, $\beta$ approaches 90°, and vector N approaches the x axis.

Although the component waves are still hyperconical, their shape also changes with wavicle velocity. In Fig. 6-6(c) the line segment F represents a section of a component wavefront, while $\alpha$ is its angle with respect to the hypercone axis and vector N. While N represents the motion of a node associated with F, the perpendicular vector, c, indicates the movement of this component wavefront section. As shown, $\alpha$ is given by

$$\sin\alpha = c/N \qquad (6\text{-}25)$$

At large wavicle velocities, where N diminishes toward c, the quantity on the right again goes to 1, and $\alpha$ to 90°. Consequently, the hypercones open out, and approach hyperplanes in the limit. So, at velocities very close to the speed of light, the component fronts approximate hyperplane waves, traveling in the x direction, with velocity c.

(Further, the incident and reflected portions of a single wavefront no longer have the same axis, although both axes parallel N. Projecting both hypercones inside the core, the vertex of each is at the core center. However the two vertices are offset in the $\phi$ dimension, so the two axes are displaced from each other, as shown in Fig. 6-6(a). Speaking in terms of the three dimensions of the figure, the core node where the two [hyper]cones join is an ellipse. This shape also characterizes the nodes in general.)

*In contrast to the component waves, which remain hyperconical, Moiré wavefronts are essentially hyperplanar.* As shown in two dimensions in Fig. 6-4, in a stationary wavicle, these are perpendicular to the ϕ axis. For a wavicle moving steadily in the x reference space dimension, the hyperplanes are tilted in x and ϕ, as illustrated in Fig. 6-5. Also, from our relativistic transform, there is no tilting with respect to y or z, so the resulting hyperplanes remain perpendicular to the x,ϕ plane.

The basic shape difference between component and Moiré wavefronts is a crucial one. *Since the Moiré fronts of a uniformly moving wavicle are hyperplanar, these appear as infinite plane waves in the reference space, exactly like the de Broglie waves of a uniformly moving particle. It follows that, knowing only the spatial arrangement of its Moiré wavefronts, one can infer the momentum of such a wavicle, but not its position.*

## Flow Wave Effects and the Wavicle Core

As described, a stationary wavicle has a flow wave pattern with an amplitude and average ϕ flow varying as $1/r^2$. Since the wave pattern of a moving wavicle contracts relativistically, its flow pattern must do this also. In addition, its average (fixed-position) flow wave frequency increases relativistically. Given that a wavicle's ϕ flow is proportional to both wave amplitude and frequency, total ϕ flow is then conserved. This matches the conservation of charge in an actual particle, where contracted (ellipsoidal) equi-potential surfaces enclose increased potentials.

For a wavicle in an external ϕ flow (corresponding to a charged particle in a potential), its wave pattern can be expected to move with the effective ϕ flow, as described previously for pro-particles. As depicted in Fig. 6-3, we have component waves of both relative densities, moving in oppositely with respect to ϕ. Consequently, half are augmented in frequency, while the other half are decreased by an equal amount. This gives an average frequency and associated ϕ flow which are unchanged. I.e., net ϕ flow is still conserved, just as charge is for an actual particle in an external potential. *Thus we have a complete model of a relativistic, quantized charge.*

Again, for a given wavicle pattern having a specific rest frequency, an inverted pattern with an opposite ϕ flow is also possible. As you might expect, *the two cases are taken to represent corresponding particles and antiparticles.* Of course wavicles with positive ϕ flows are the ones we'll choose to equate with positively charged particles.

In the simplified wavefront diagrams above, we've assumed that a wavicle's components move at a constant velocity, c, and thus can retain a hyperconical shape. Clearly this is reasonable at large distances from the core, where the amplitudes and nonlinear effects are vanishing. What happens in the region close to the core (this may be relatively small), where the flow wave amplitudes are great? Here the picture is less clear.

For individual component waves taken in isolation, it may be that their speed would vary with amplitude. (Like familiar shock waves, the waves with greater density might move faster.) Taken together, we also expect a wavicle's components to simultaneously reflect, refract, and displace one another. In a balanced system, where the overall density remains constant, some of these nonlinear effects may tend to cancel. So it's conceivable that the effective component wave velocities might remain approximately c near the core.

(This would imply that the flow wave interactions within wavicles differ from those between wavicles. Similar behavior is observed in solitons, where the component wave interactions differ from those of whole solitons.)

Whatever transpires near the wavicle center, it's required that, after leaving this region, the various outgoing component waves should be the reflected equivalent of those incoming. (This is necessary to conserve ϕ flow, density, and energy together.) *Since wavicles are taken to be self-organizing, there should also be feedback in the core region, actively adjusting the waves to fit this pattern.* It's hoped that such behavior can be verified by modeling wavicles numerically.

And the core? The example of acoustic cavitation suggests the possibility of a distinct surface, representing a sharp discontinuity in the properties of the wave medium outside and inside the core. (To simulate such a wavicle numerically, it would be necessary to model a separate set of conditions existing inside the core.) With an overall alignment paralleling the ϕ axis, the surface conceivably might be a tubular one, rippled by impinging waves. Or, it could consist of separate bubbles, in a periodic string with the same orientation.

To describe the possible interactions of wavicles, we can begin by considering the responses of individual wavicles to external ϕ flows. Along the lines of our earlier work with pro-particles, we can approach this from the standpoint of an individual wavicle encountering a "sharp" change in ϕ flow (electrical potential). However, here all external ϕ flows are to be attributed to other wavicles. So in this case, the external flow wave pattern is necessarily complex, with the actual ϕ flow at a given instant depending on the positions, phases, etc. of the various other wavicles.

Earlier, we saw that the component waves in this system are simultaneously reflected and transmitted at a sharp ϕ flow interface. However, for a wavicle to persist, the overall system must "decide" whether the whole entity is to be transmitted or reflected. The system's behavior must be such that, after the wavicle's motion changes, waves originating from prior times continue to arrive at its future core position. To conserve the total momentum of the system, other wavicles necessarily undergo opposing changes of motion. And these also must continue to receive incoming waves.

Similar behavior is found in other nonlinear wave systems, such as that of Zabusky and Kruskal,[8] cited in Section 1. Each of its various solitons can be viewed as a collection of Fourier component waves, shared with the others in the system. As the solitons interact and their individual velocities change, all the various components are observed to change with them, accordingly. Thus each hump retains its due share of component waves and persists indefinitely. (Since they interact and pass through one another "without losing their form or identity," wavicles could possibly be classified as solitons. However, the term normally refers to running waves, rather than standing wave patterns.)

From our earlier pro-particle work, a free wavicle's velocity can be expected to change if it enters a different ϕ flow. Also, we've seen that the wave pattern depends on its velocity. So as the wavicle core crosses the interface, its pattern must change -- initially for the impinging waves close to the core. This presumably occurs via a nonlinear wave interaction -- one which simultaneously alters the patterns of other wavicles in the system. The region of nonlinear interaction should then expand outward.

In this theory, *such propagating disturbances in the patterns of wavicles represent electromagnetic waves.* (Here electromagnetic waves alone can't exist. In "empty" space, they represent disturbances in the wave fields of distant matter.)

As described, the Moiré waves of wavicles are taken to correspond to de Broglie waves. Note that a wavicle has two sets of Moiré waves -- one associated with its positive density components, and another with those of negative relative density. The wavelengths of the two sets in the reference space are always identical. However, in external ϕ flows their frequencies differ.

Again, for actual charged particles in scalar potentials, the de Broglie frequency and energy (or Hamiltonian) are related by

$$h\nu = E = \frac{m_0 c^2}{\sqrt{1 - v^2/c^2}} + q\Phi \qquad (6\text{-}26)$$

So, where the charge, q, and scalar potential, Φ, are of the same sign, the de Broglie frequency, ν, is increased. For a wavicle in a ϕ flow of the same sign, its positive Moiré waves are increased in frequency, analogously. Meanwhile, the frequency of its negative ones is decreased.

Thus, if wavicles are to behave like actual particles, it's their *positive* density Moiré waves which must correspond to de Broglie waves. For example, if a wavicle enters a stationary ϕ flow region, its velocity should change such that the positive Moiré frequency is conserved. It may be difficult to show directly that wavicles act this way. However, it's reasonable that they would, since the total energy, momentum and ϕ flow of the system are conserved if they do. Again, it may be possible to verify this through computer simulation.

Given that our system behaves this way, we have entities which both generate and respond to potentials as actual particles do. And, based on different characteristic wave angles, particles of different masses can be modeled. Tentatively, we'll take individual wavicles to represent the charged leptons: the electron, muon, tauon and their antiparticles. Since there is no evidence of internal structure in these, single wavicles with small cores may be an accurate representation. Again, this avoids the problem of infinite fields encountered in standard field theory,* where these particles are treated as point charges.

---

*To this end, a very similar model of electrons and muons was proposed by Dirac in 1962.[32] In it, the electron is viewed as "a bubble in the electromagnetic field", of finite size. The muon is then taken to be an excited state of the electron, involving radial pulsations of the bubble and a spherically symmetrical wave field.

# 7. Future Additions

Several additional sections of this paper are planned, but incomplete at this point. Here I'd like to try to give some sense of what these are about and to introduce some of the key ideas. (At my usual glacial pace, it will be some time before the remaining sections are written. So it seems best to go ahead, publish this much, and finish later!) Also, the concluding section, which follows this one, is more-or-less complete.

## Quantum Mechanics

One added section will address quantum mechanics, with particular attention to the Einstein-Podolsky-Rosen, or EPR, effect. The starting point is the observation that, for a wavicle to persist, even its simplest behaviors require the coordinated action of the overall wave system. Thus a wavicle's actions are necessarily determined by non-local factors. Such non-locality is also characteristic of the EPR effect.

It's sometimes said that each part of a nonlinear wave system "knows" what the other parts are doing. The quasicrystalline wave structures mentioned in Section 1 are a clear example. (As Penrose[33] points out, the formation of a quasicrystalline patterns requires knowledge of remote parts of a system.) Connected, non-local wave phenomena of this sort can be seen as the result of nonlinear coupling. (Like that of nonlinearly coupled pendulums or oscillators.) Such coupling is expected with the pulsations of wavicles, and is taken as the basis for quantized de Broglie wavelengths, such as those of electrons in atoms.

Again, the present work is related to Bohm's "hidden variables" theory,[2] where particles exist continuously in definite states and follow definite trajectories. In it, de Broglie waves and particles are influenced by an additional "quantum-mechanical" potential or "quantum potential," which has a non-local aspect. Bohm's theory is based on the non-relativistic Schrödinger wave equation

$$i\hbar \frac{\partial \psi}{\partial t} = -\frac{\hbar^2}{2m}\nabla^2 \psi + V\psi \qquad (7\text{-}1)$$

His first step was to rewrite the wave function ψ as

$$\psi = R e^{iS/\hbar} \qquad (7\text{-}2)$$

where R and S are real, and represent its amplitude and phase respectively. Putting this simple wave description into Schrödinger's equation, Bohm derived this one:

$$-\frac{\partial S}{\partial t} = \frac{(\nabla S)^2}{2m} + V - \frac{\hbar^2}{2m}\frac{\nabla^2 R}{R} \qquad (7\text{-}3)$$

The term on the right corresponds to the quantum potential, which he expressed as

$$U = -\frac{\hbar^2}{2m}\frac{\nabla^2 R}{R} \qquad (7\text{-}4)$$

Bohm also assumed that particle velocity vector, **v**, is given by

$$\boldsymbol{v} = \nabla S / m \qquad (7\text{-}5)$$

He then showed that, if the quantum potential were to fluctuate randomly, such particles would behave exactly in accordance with the usual statistical laws of quantum mechanics -- despite following well-defined trajectories. (Striking computer-modeled trajectories of particles in quantum potentials can be found in Philippidis et al.[34] or Vigier et al.[3]) Bohm and Vigier[2] subsequently proposed random fluctuations in a fluid wave medium as a physical basis for the quantum potential.

Toward a relativistic theory, Vigier[3] has since proposed a "sub-quantum Dirac ether" as the fluid medium. Unlike real fluids, the elements of a Dirac ether have no specific states. (Dirac[35] describes it as analogous to the concept of particle clouds, where particles lack definite states.) "Manifest" relativistic covariance then is permitted by the fact that there is no overall state of motion. However, this is inconsistent with the basic de Broglie/Bohm theory, where particles have definite states and the wave medium is taken to be physically real. On the other hand, the present theory offers an appropriate medium which is both relativistic and realistic.

At this point, we'll introduce another feature of wavicles. In the previous section, their wavefronts were illustrated as perfectly symmetrical hypercones, for simplicity. However, in a realistic case, where the source of a wavicle's incident waves is others, the patterns are necessarily irregular to some degree. (There is no way to generate perfectly symmetrical hyperconical wavefronts from sources at discrete, scattered sites.) While feedback maintains a wavicle's average ϕ flow, such asymmetries would result in intermittent fluctuations in the ϕ flow seen at the core, and changes in its motion. Chaotic ϕ flow fluctuations of this nature are taken to correspond to Bohm's quantum potential.

# Gravitation and Cosmology

In Einstein's version of general relativity, gravitation is based on a varying non-Euclidean space-time, and an absolute speed of light. In this theory, gravity is associated with local variations in the speed of light, while space and time (in a preferred frame) are absolutes. For very strong gravitational fields, or large astronomical distances, the two theories make very different, testable predictions.

Like the electromagnetic forces, the source of gravitation here lies in the wave fields of wavicles. In concentrations of matter containing oppositely charged wavicles, the average ϕ flows tend to cancel at large distances, like the electric fields of conventional particles. However, *the underlying wave fields do not cancel.* The cumulative waves fluctuate more-or-less randomly, with an rms amplitude varying approximately as $1/r^2$ from the center of mass, like the wave fields of individual wavicles. The resulting irregular fluctuations of the wave medium are taken to represent gravitational potentials. (Thus these and the quantum potential are related.)

The key to gravitation in this system is the dependance of the rate of physical processes on the 4-D wave speed, c. Our wave-based gravitational potentials act by reducing this general velocity. (This may be attributable to disruption of rectilinear wave propagation by the medium's fluctuations. For example, a ray traveling from point a to point b experiences small-scale excursions which increase the total path length, effectively decreasing the wave velocity.) As c is reduced, everything is slowed proportionately -- the movements of wavicles, the propagation of electromagnetic waves, etc.

Clocks in gravitational potentials are perceived by outside observers to be slowed, as usual. Light is also bent in a gravitational field. In this case, though, the effect corresponds to ordinary refraction. (Of course refractive index goes as the inverse of wave speed.) As a result, the local behavior of physical systems in gravitational fields matches that in accelerating frames -- i.e., Einstein's principle of equivalence is met.

For comparing Einstein's account of gravity with this one, the concept of optical path length is useful. Often used in evaluating optical systems, this is computed by multiplying the lengths of light rays in optical elements by the refractive indices there. While treated as distances, these products are actually a measure of the time required for a wavefront to transit an element. (In an ideal imaging system, the summed optical path lengths between corresponding object and image points are the same for all rays. This says that all parts of a wavefront arrive at an image point simultaneously -- in phase.) Still, in terms of light wavelengths, a material with an index above one behaves as though additional space is compressed inside it.

Einstein assumes that optical path lengths in gravitational potentials are the same as spatial distance.  Here, as in ordinary optical systems, they aren't.  As a result, different arrangements of things in space are predicted.  For weak gravitational fields and modest astronomical distances, like those within the solar system, the differences are minute.  At cosmological scales, however, these should be apparent.

Because c isn't sacrosanct in this theory, an alternate basis for the Hubble redshift is also permitted.  Here, c is taken to be gradually increasing.  Since this corresponds to the rate of all processes (at the macroscopic level), at the time of emission, the spectra of remote galaxies would have been lower in frequency than present ones.  So we're again invoking a varying wave speed, instead of a varying space-time.

To provide for a changing c, we'll take the wave medium, stuff, to resemble a real gas, with particle-like constituents.  The characteristic acoustic wave velocity then depends on the effective temperature and density.  We can also suppose that the constituents of stuff are also wave-based entities (like tiny wavicles), associated with some finer coherent structure.  If energy is gradually transferred between wave structures at different scales, we then have a mechanism for changes in these qualities of stuff at the macroscopic level.

(Stuff is assumed to have a fractal-like nature, with wave structures arising and subsiding at widely differing scales.  In this respect there's a resemblance to "inflationary" cosmological models.)

Such energy transfers occur naturally in nonlinear wave systems.  Not only from large waves to small, but also in the opposite direction.*  An important example was discovered in computer-modeled waves by Fermi, Pasta and Ulam in the 1950's.[10]  Prior to their work, it was widely believed that nonlinearity in waves inevitably leads to thermodynamic behavior, where orderly large-scale waves degenerate into random small-scale ones.  The system studied was one-dimensional, with a discrete length.  Initially, a large sinusoidal wave was seen to transform into various configurations of smaller ones.  However, the process eventually reversed, with the small waves coalescing back into the starting waveform.  (The effect is now called FPU recurrence.  Fermi, Pasta, and Ulam also inspired the computer experiments of Zabusky and Kruskal, discussed earlier.)

It seems clear that general qualities of the universe do undergo long-term changes.  In the course of stellar evolution, a substantial fraction of the collective mass in stars is being converted to neutrinos.  However, the bulk of these are undetected and unaccounted for.  It's conceivable that much of the lost energy is transferred to a smaller-scale wave structure, helping raise the effective "temperature" of stuff at that level, and the value of c.

---

*Large ocean waves are now thought to grow from smaller ones overtaken.[7]

According to Einstein's general relativity and the standard Big Bang model, a curved, finite, expanding space-time is the basis of the Hubble redshift. On the other hand, *the repeating wave system in this theory requires a Euclidean space.* Since different topologies are predicted, the two theories can be tested by determining the structure of space. Of course any direct observation of space-time curvature (including entities such as wormholes) would invalidate the new theory.

The large-scale curvature of space can be estimated (at least to an order of magnitude) by counting galaxies within different radii from our position. In a flat space, the numbers would increase approximately as the cube of the radius, while a curved space-time should deviate from this. Studies of this kind, for example by Loh and Spillar[36], have indicated that space is either flat, or very close to it. According to Linde[37], the discrepancy between the observed curvature and that predicted by Einstein's general relativity and the standard Big Bang model is roughly *sixty* orders of magnitude. This extreme disagreement is called the "flatness problem."

While the present theory predicts flatness, it may seem trivial, since a Euclidean space was assumed at the outset.* However, note that this assumption is an essential element of the basic theory, and not something added to fit certain astronomical data. In comparison, the assumptions regarding space are much more complex in Einstein's theory. There space is Minkowskian, also curved by the presence of matter, and possibly further reshaped by an additional cosmological constant.** Of course the overriding point is that, unlike the standard Big Bang model, here space-time curvature is *not required*.

Recent observations pertaining to the age of the universe also appear to contradict the Big Bang. Attributing the redshifts of remote galaxies to recessional velocities, the predicted age can be obtained by dividing the distances of galaxies by their apparent velocities. Using the Hubble Space Telescope to estimate the distance of the galaxy M100, Freedman et al.[38] have put the universe's age at approximately 8 billion years -- less than the estimated age of its oldest stars.

The discovery of highly-developed large-scale structures, such as the Great Wall and Great Attractor, also argues for an older universe than that allowed by the standard Big Bang model. In this theory, the age of the universe isn't similarly constrained, since the Hubble redshift is not attributed to an expanding universe. Ancient stars are permitted, and mature large-scale structures are expected.

---

*Poincaré argued that the true geometry of the universe *must* be Euclidean.

**The cosmological constant, $\Lambda$, from Einstein's early gravitational field equations, has been proposed as a means of dealing with the flatness problem. Still, even with the re-introduction of this extra, arbitrary quantity, a sufficiently flat space appears unattainable within the constraints of standard Big Bang cosmology.[37]

# Nuclear Forces and Sub-particles

This part of the theory is more tentative; at this point it's only qualitative.

Presently, nuclear forces are taken to be relatives of the Casimir effect. The latter involves vacuum fluctuations (random electromagnetic waves occurring in "empty" space) which are suppressed in regions between parallel conducting plates. As a result, waves are reflected disproportionately from the outside plate surfaces, creating a small pressure driving the plates together. The force varies as the inverse of the plate spacing, to the fourth power.

Due to their finite sizes, similar effects may arise between reflecting wavicle cores. Although these are not flat plates, in close proximity, they would shield one another somewhat from incident component waves. Also, since the component wave amplitudes are extreme near the cores, a large attractive force could arise at short distances. This effect, which depends in part on the wavicle core sizes and the characteristic hyperconical wave angles, is the assumed basis of nuclear forces. (The feedback which sustains the individual wave patterns is also taken to oppose the complete merging of wavicles.)

Since multiple nonlinear effects are involved, it appears difficult to test this idea by analytical methods. However, it may be possible to verify it through computer simulation of the wave system. (A possibly related effect is seen in ultrasonic wave tanks, where the cavitating fluid contains particulate impurities. The wave action tends to drive these together, forming aggregates which precipitate out. The process is sometimes used to purify fluids. Suspended particles of metal powder can be driven together so forcefully that they partially melt and fuse.[39])

Besides constituting nuclei, bound groupings of wavicles may also be needed to account for the internal structure of hadrons. If so, the binding forces might be similar to those of nuclei. In the prevailing quark model, the constituents of hadrons have fractional charges. On the other hand, in the current version of this theory, uniform (or nearly uniform) elementary charges are called for. A compromise might be to assume that both hadrons and leptons have constituents with charges of ±⅓e. A model of this nature has been proposed by Harari[40], for example.

Still, fractional charges have never been observed directly. Also, the prevailing quark model is contradicted by recent measurements of the proton's spin magnetic moment, which give only about one third the predicted value.[41] (The "spin crisis.") Since charge and spin magnetic moment are related, one wonders whether this might indicate that hadrons actually have constituents with integral charges. As described by Nambu and Han[42], various models based on quark-like sub-particles with integral charges have been suggested in the past.

# Computer Modeling

Again, some wave behaviors required by this theory are difficult to verify analytically. (The evolution of nonlinear waves in a gas-like medium can be viewed as an extreme example of the many-body problem -- with each particulate constituent a separate body.) However, it may be possible to do this though computer modeling. (No modeling has been attempted at this point.) Some basic questions to be answered are:

> 1.) Can a wave system of this kind remain balanced? (Complimentary positive and negative density flow waves, balanced such that net flows only arise in the $\phi$ dimension.)
>
> 2.) To arrive at the vector potential, we assumed that the effective $\phi$ flow experienced by pro-particles at their moving positions is proportional to the frequency of flow waves encountered. Does this hold?
>
> 3.) Are self-organizing flow wave patterns resembling wavicles observed?
>
> 4.) Do these exhibit quantum-mechanical behaviors?

As discussed in Section 1, the computer experiments of Zabusky and Kruskal modeled an infinite one-dimensional system of nonlinear waves. Because these were periodic, it was only necessary to model a single region equal in length to the period. Waves were wrapped around from one edge of the region to the other. With respect to the $\phi$ dimension, we can do the same for the periodic waves of this system.

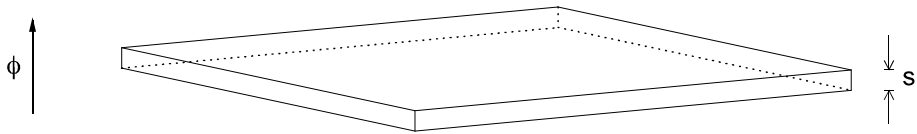

**Fig. 7-1.** 3-D representation of a rectangular hyperbox-shaped region for modeling the evolution of waves in this system. Depicted are $\phi$, plus two arbitrary reference space dimensions.

While only a single layer of the system needs representation in the $\phi$ dimension, a relatively broad region needs to be covered in x, y, and z. A simple choice of region to model would be a thin, rectangular hyperbox, illustrated in three dimensions in Fig. 7-1. The box height, s, is the $\phi$ period of the wave structure.

Waves at the top and bottom of the hyperbox would wrap around. To prevent waves from escaping, the sides could be treated as mirrors. Of course standing waves in a cavity are strongly influenced its shape. (A contained wavicle might be seen to behave like a quantum-mechanical particle in a box.) A hyperbox with an irregular footprint in the reference space might give a better approximation of waves evolving in an infinite space.

Another possible way to maintain the waves would be to also wrap them from side wall to opposite side wall, instead of reflecting them. Further strategies could be used to reduce the regularity of the "cavity." For example, different sections of a given side could be wrapped to sections of various other sides. Instead of wrapping directly across, opposite sides could be wrapped with a twist. Wrapping and reflection could be combined. (Reflection is essentially wrapping a side to itself.) Etc . . .

The medium, stuff, could be represented either as a continuous fluid, or one with discrete, particle-like constituents. For a realistic simulation in four dimensions, the required number of particles or fluid elements is huge -- almost certainly more than $10^{11}$. (Enough to challenge a massively parallel supercomputer.[43]) Modeling an analogous three-dimensional system would cut the computational demands greatly.

Something similar to wavicles, with conical rather than hyperconical wavefronts, might arise in 3-D systems. However, if such patterns occur, their associated $\phi$ flows would vary as $1/r$, rather than $1/r^2$, giving them very different dynamical properties. Because wavicles depend for their existence on interactions with others, lacking the same dynamics, their analogues may not form. Even so, basic questions about the behavior of 4-D wave systems should still be answerable from three-dimensional experiments. (For example, questions 1 and 2 above.)

Probably the biggest obstacle to constructing an accurate model is the fact that the specific characteristics of stuff are unknown at this point. Like real fluids, the behavior of stuff may change abruptly at critical points. From the example of ultrasonic cavitation bubbles in pure water, stuff inside a wavicle core may have sharply different properties.\* A possible approach to modeling wavicle formation would be to take an arbitrary set of balanced flow waves, and "seed" it with hyperbubbles, aligned with the $\phi$ axis. (The stuff inside these could be assigned a much higher wave velocity, for example.) Waves might nucleate around these, forming wavicle-type patterns, with hyperbubbles as cores.

Zabusky and Kruskal's modeling of a one-dimensional, repeating wave system was very illuminating. In higher-dimensioned systems like this one, it seems likely that richer forms of self-organization will be found. Whatever the wave behaviors may be, it's certain they'll be interesting.

---

\*Even knowing something of the properties of normal water, we don't yet have a good model for the waves surrounding acoustic cavitation bubbles.

# 8. The Course of Physics

All of modern physics is built on Einstein's interpretation of special relativity. Included is his view that, a priori, systems with preferred frames, like Lorentz's, must be ruled out. (In contrast, Lorentz was willing to concede the possible validity of Einstein's space-time approach.) In the early years, this exclusionary position was criticized by leading theorists. Sommerfeld, for example, called it "unhealthy dogmatism." However, after the apparent confirmation of Einstein's general relativity, his views were adopted by the physics community at large.

The various founders of quantum mechanics unanimously took space-time as its framework. In his stand against the ether, Einstein had argued we should not speak of things that can't be measured. Heisenberg, Bohr, and others extended this philosophy also to quantum mechanics. Their Copenhagen interpretation says that, for the same reason, we shouldn't speak of hidden variables. This viewpoint found its ultimate expression in Bohr's statement that the quantum mechanical microworld "doesn't exist."

As we know, Einstein was dismayed by this, and took a lonely (and courageous) stand in defense of the principle of causality. Eventually, a talk with Einstein turned Bohm against the Copenhagen interpretation, and spurred the development of his hidden variables theory. Bohm's success also inspired de Broglie to resurrect his related "pilot wave" theory. Neither was endorsed by Einstein, however, who saw a conflict between his own conception of relativity, and the non-local effects described by Bohm.

Both de Broglie and Bohm were devout believers in Einstein and Minkowski's space-time. However, Bell has noted that their theory appears to require a preferred frame of reference. He writes:

> I think that conventional formulations of quantum theory, and of quantum field theory in particular, are unprofessionally vague and ambiguous. Professional physicists ought to be able to do better. Bohm has shown us a way. It will be seen that all the essential results of ordinary quantum field theory are recovered. But it will be seen also that the very sharpness of the reformulation brings into focus some awkward questions. The construction of the scheme is not at all unique. And Lorentz invariance plays a strange, perhaps incredible role . . .
>
> ...I am unable to prove, or even formulate clearly, the proposition that a sharp formulation of quantum field theory, such as that set out here, must disrespect serious Lorentz invariance. But it seems to me that this is probably so. As with relativity before Einstein, there is then a preferred frame in the formulation of the theory . . . but it is experimentally indistinguishable. It seems an eccentric way to make a world.[21]

Bell further points out the legitimacy of a preferred reference frame in "How to teach special relativity"[15], where he advocates the approach of Lorentz and Poincaré.

In his long search for a unified theory, Einstein followed a belief in the underlying simplicity of physical phenomena. He said "If the answer is simple, God is talking." He also looked to geometry as the source of this simplicity, believing that natural phenomena should be describable in geometric terms. His early paper, "On the Electrodynamics of Moving Bodies," showed that, in a space-time representation, electric and magnetic fields are exactly the same. Einstein took this unification as proof of space-time, and an essential step toward a unified field theory. Still, space-time geometry has not delivered such a theory.

In accordance with Einstein, it is now generally held that proper mathematical descriptions of relativistic systems should be "manifestly covariant." That is, they should make no reference to any sort of preferred frame. However, even in classical mechanics, this leads to difficulties. According to Dirac[44], we now know that "*the Hamiltonian form for the equations of motion is all important*." (His italics.) On this representation of mechanics, Goldstein's graduate text[19] observes:

> As with the Lagrangian picture in special relativity, two attitudes can be taken to the Hamiltonian formulation of relativistic mechanics. The first makes no pretense at a covariant description but instead works in some specific Lorentz or inertial frame. Time as measured in the particular Lorentz frame is then not treated on a common basis with other coordinates but serves, as in nonrelativistic mechanics, as a parameter describing the evolution of the system. Nonetheless, if the Lagrangian that leads to the Hamiltonian is itself based on a relativistically invariant physical theory, e.g., Maxwell's equations and the Lorentz force, then the resultant Hamiltonian picture will be relativistically correct. The second approach, of course, attempts a fully covariant description of the Hamiltonian picture, but the difficulties that plagued the corresponding Lagrangian approach are even fiercer here.
>
> ...there seems to be a natural route available for constructing a relativistically covariant Hamiltonian. But the route turns out to be mined with booby traps.

Goldstein also shows that the preferred frame method easily gives a concise, non-covariant Hamiltonian (Eq. (5-16) in this paper). Applying Einstein's own dictum about the simple answer, in this case it says that manifest covariance is *wrong*.

As mentioned earlier, space-time suffers other drawbacks in comparison with the preferred frame approach. One is its inconsistent treatments of linear and rotational motions. Although forbidden for linear motion, a preferred frame defined by "the fixed stars" remains for the rotational case. So, instead of a single reference frame, there are two very different types.

Again, an egregious shortcoming is that space-time by itself can't account for the direction of entropy. Since its time dimension must be exchangeable with its spatial ones, time necessarily has a bidirectional character, with no favored direction.* Another parameter, like the unidirectional, independent time of the preferred frame approach, is required. Thus Minkowski space-time avoids neither a preferred spatial frame, nor apparently an independent time -- this despite its added geometric complexity.

In the present theory, Einstein's space-time view is abandoned. However, several of his most basic ideas remain: That the universe is fundamentally simple and describable in terms of a special geometry. His requirement of causality. And the same relativistic principles are followed -- both Poincaré's principle of relativity, and Einstein's principle of equivalence. Some other key ideas of this theory are these:

> 1.) Contrary to the prevailing view, transverse wave phenomena *can* arise in continuous fluid media. Where there are nonlinear wave structures, transverse waves of interaction can arise *within* the waves. Such waves of interaction are taken to represent electromagnetic ones.
> 
> 2.) The unlimited velocity of matter waves is directly explainable on the basis of a repetitive four-dimensional wave structure. Its geometry is Euclidean, and the waves are redundant in one dimension, $\phi$. (Consequently, this dimension isn't obvious to observers.) When the wave movement parallels $\phi$, the apparent 3-D velocity is infinite.
> 
> 3.) The wave medium is an elastic fluid. Hence its waves are acoustic and have a characteristic 4-D velocity, c, under standard conditions. Moving wave-based reflectors, aligned with the $\phi$ axis, transform the waves relativistically.
> 
> 4.) The scalar potential can be exactly accounted for by flows in the $\phi$ dimension, from their refractive and reflective effects on matter waves.
> 
> 5.) These flows are driven by flow waves, balanced such that net flows only arise in the $\phi$ dimension. This preserves the relativistic character of the system, and also accounts for the vector potential.
> 
> 6.) Like other nonlinear waves, those of this system are self-organizing, forming additional patterns, called "wavicles," within the overall wave structure. These behave as elementary particles. Involved are approximately hyperconical waves, reflected at a central "core" region where the amplitudes are great. (Wavicles are also the above wave-based reflectors.)

---

*It's sometimes argued that this problem can be resolved by assuming that, when the time dimension of space-time is reversed, entropy is too. The reversed case would be perceived as normal by observers, since their thought processes would run backwards also. However, since space-time is symmetrical, this still provides no basis for the acknowledged difference between the forward and reverse cases.

7.) A wavicle's components are flow waves. The resulting $\phi$ flow pattern corresponds to the electromagnetic potentials of a charged particle. Noninverted and inverted patterns represent corresponding particles and antiparticles. Quantized charge and energy result from a characteristic limit to the $\phi$ flow attainable at the core.

8.) Wavicles are inherently non-local phenomena, interdependent on other wavicles and their reflected waves for their existence. Their behavior is that of extended objects, of arbitrary extent. Like solitons, wavicles move and interact as wholes.

9.) The synchronous pulsation of a wavicle's standing wave field can be characterized in terms of "Moiré wavefronts," representing surfaces of equal phase. These correspond to de Broglie waves, and define the quantum mechanical behavior of wavicles. Nonlinear coupling of the pulsations, both within and between wavicles, is taken as the underlying mechanism.

10.) In concentrations of matter, the average $\phi$ flows of oppositely charged wavicles tend to cancel at large distances, like the electric fields of conventional charged particles. The underlying waves do not cancel, however. Consequently, intermittent flows arise, with an rms amplitude varying, like the individual wave fields, as $1/r^2$ from the center of mass. These intermittent flows decrease the effective velocity of waves; i.e., the value of c is locally reduced. This slowing is taken as the basis of gravitation. Since the 4-D velocity, c, determines the rate of all physical processes, these are observed to slow also, as in Einstein's general relativity. In this case, the gravitational bending of light corresponds to ordinary refraction.

11.) The wave medium, stuff, is assumed to have a fractal-like character, with additional wave structures at much larger and smaller scales. As in FPU recurrence, wave energy can be gradually transferred up or down between scales, changing the effective temperature or density of stuff at those levels. Because acoustic wave velocity depends on these parameters, c is also allowed to change over time. (Again, this determines the rate of physical processes at the macroscopic scale.) An increasing c is proposed as the cause of the Hubble redshift seen in light from distant galaxies.

Until recently, the Hubble redshift has been taken as final proof of Einstein's general relativity. The effect was thought to be readily explained on the basis of a finite, expanding universe, allowed by a curved Minkowski space-time. However, the original Big Bang model is contradicted now by a variety of observations. An example cited previously is the distance to the galaxy M100 measured by the Hubble Space Telescope. Based on the standard Big Bang model, this puts the age of the universe at about 8 billion years. This is under the estimated age of the oldest stars.[38]

It's also difficult to reconcile the Big Bang with the discovery of very-large-scale structures, such as the Great Wall or the Great Attractor. Most importantly, there is the flatness problem. From the observed distribution of galaxies, the large-scale curvature of the universe is relatively small, or nonexistent. Again, according to Linde[37], the discrepancy between the observed curvature and that predicted by general relativity and the Big Bang is roughly *sixty* orders of magnitude. Astronomical observation, once taken as confirmation of Einstein's general relativity, now speaks against it.

In contrast, the theory presented here allows a macroscopic universe older than its stars, with mature structures, and predicts an inherently flat geometry. (In addition, it predicts new nonlinear wave effects, which may be verifiable through computer simulation.) There is also the broad unification of physics, in a much simpler paradigm. For the first time, relativity and quantum mechanics are joined. Both the wave and particle aspects of elementary particles find a single, coherent representation. And there appears to be a truly unified physical basis for the known natural forces.

Einstein argued that a preferred frame should not be hypothesized if it can't be identified experimentally. However, at least a very special case of linear motion is precisely defined by the cosmic microwave background. This apparently corresponds to the overall system of celestial matter -- the same one Einstein took as the preferred frame for rotation. Since there presumably is some coupling between its different scales, it's no less reasonable to take this as the frame of our wave medium, stuff.

Physics has seen so little fundamental progress, for so long, it's often remarked that "physics is dead." Is it proper, then, to continue taking our most fundamental physical assumptions on *faith*? Shouldn't we have the courage to explore valid alternatives, such as the relativity of Lorentz and Poincaré, and the quantum mechanics of de Broglie, Bohm and Vigier? To solve a maze, sometimes it's necessary to double back and take a different path.

# Acknowledgments

Thanks to Eduardo Cantoral for his counsel. And I'd especially like to thank Surrendar Jeyadev, who edited the manuscript. (The remaining informal language is all my fault.) This journey of ideas is dedicated Ken and Pat Krogh, who encouraged every step.